\documentclass[lettersize,journal]{IEEEtran}
\usepackage{amsmath,amsfonts}
\usepackage{array}
\usepackage[caption=false,font=normalsize,labelfont=sf,textfont=sf]{subfig}
\usepackage{textcomp}
\usepackage{stfloats}
\usepackage{url}
\usepackage{verbatim}
\usepackage{graphicx}
\usepackage{cite}
\usepackage{bbm}
\usepackage{subfig}
\usepackage{amsthm} %required for proof env
\usepackage{amsmath} %required for multiline equation
\newtheorem{theorem}{Theorem}

\usepackage[linesnumbered,ruled,vlined]{algorithm2e}
\SetKwInput{KwData}{Input}
\SetKwInput{KwResult}{Output}

\usepackage{mathtools}
\newtheorem{lemma}[theorem]{Lemma}

\newcommand{\myDS}{interaction\_matrix}

\usepackage{amsfonts}%added to support mathbb

\usepackage{orcidlink}

\begin{document}

% The next 4 commands, are absolutely "Non plus ultra"
% \setlength{\abovedisplayskip}{3pt}
% \setlength{\belowdisplayskip}{3pt}
% \setlength{\textfloatsep}{3pt plus 1.0pt minus 2.0pt}
% \setlength{\intextsep}{3pt plus 1.0pt minus 2.0pt}

\title{A Parallel Algorithm for Finding Robust Spanners in Large Social Networks}
% \title{A Parallel Algorithm for Robust Spanner Detection in Large Social Networks}

\author{
    Arindam Khanda\textsuperscript{\textdagger}\orcidlink{0000-0003-3364-8914},
    % ~\IEEEmembership{Member,~IEEE,}
    Satyaki Roy\orcidlink{0000-0001-6767-266X},
    % ~\IEEEmembership{Member,~IEEE,}
    Prithwiraj Roy\orcidlink{0000-0002-0592-5593}, and
    % ~\IEEEmembership{Member,~IEEE,}
    Sajal K. Das\orcidlink{0000-0002-9471-0868},~\IEEEmembership{Fellow,~IEEE}
\thanks{\textsuperscript{\textdagger}Portions of this manuscript are included in the first author’s Ph.D. dissertation: A.~Khanda, \textit{Parallel Algorithms for Large Scale Dynamic Graph Analysis}, Missouri University of Science and Technology, 2024.}
\thanks{Arindam Khanda, and Sajal K. Das are with the Department of Computer Science, Missouri University of Science and Technology, Rolla, MO, USA. e-mail: \{akkcm, sdas\}@mst.edu\\
Satyaki Roy is with the Department of Mathematical Sciences, University of Alabama in Huntsville, USA. e-mail:sr0215@uah.edu\\
Prithwiraj Roy is with Global Action Alliance in New Jersey, USA. e-mail: proy@globalactionalliance.net}
}
\maketitle

\begin{abstract}
Social networks, characterized by community structures, often rely on nodes called {\em structural hole spanners} to facilitate inter-community information dissemination. However, the dynamic nature of these networks, where spanner nodes may be removed, necessitates resilient methods to maintain inter-community communication. To this end, we introduce robust spanners (RS) as nodes uniquely equipped to bridge communities despite disruptions, such as node or edge removals. 
We propose a novel scoring technique to identify RS nodes and present a parallel algorithm with a CUDA implementation for efficient RS detection in large networks.
Empirical analysis of real-world social networks reveals that high-scoring nodes exhibit a spanning capacity comparable to those identified by benchmark spanner detection algorithms while offering superior robustness. Our implementation on Nvidia GPUs achieves an average speedup of $244\times$ over traditional spanner detection techniques, demonstrating its efficacy to identify RS in large social networks.
\end{abstract}


\section{Introduction}
\IEEEPARstart{C}{omplex} networks are characterized by the presence of identifiable subsets of densely connected nodes, called \textit{communities}, facilitating efficient information exchange between peers within the same group. The flow of information between communities is governed by community boundaries or bridges. \textit{Structural hole spanners} (SHS) is defined as bridge nodes that allow communication between loosely connected or disconnected communities~\cite{lou2013mining}. The ability of SHS to propel information to a broad and disparate audience makes them instrumental in a variety of social network applications such as the fields of marketing, social and political campaigns, movements, awareness initiatives, etc. They are invaluable for network designers, researchers, community organizers, content creators, and individuals seeking to enhance interaction and maximize the reach and impact of any promotional efforts. Thus, a considerable amount of research has been conducted on identifying influential SHS in static social networks~\cite{lou2013mining,tang2012inferring,wang2011detecting,katz1964personal, granovetter1973strength, li2019distributed, song2015mining, rezvani2015identifying, he2016joint}.

% {\color{red}
% Complex networks are characterized by the presence of distinct communities, causing individuals to interact predominantly with peers within their group. This necessitates the identification of \textit{structural hole spanners} (SHS), which are defined as connecting nodes that allow communication between otherwise disconnected communities~\cite{lou2013mining}. The ability of SHS to propel information to a broad and disparate audience makes them instrumental in the realm of marketing, social and political campaigns, movements, or awareness initiatives that social networks are leveraged for. They are invaluable for network designers, researchers, community organizers, content creators, and individuals seeking to enhance interaction and maximize the reach and impact of any promotional efforts.

However, real-world networks are dynamic in nature, with individuals and their social connections continually forming and dissolving over time~\cite{doreian2013evolution,barabasi2002evolution}.
In such networks, spanner nodes play an even more vital role, as their impact may significantly decline if they are detected and disconnected by disrupting their inter-community links.
In such scenarios, the ability of spanner nodes to uphold network resilience and bridge structural holes is undermined, as they are susceptible to targeted link disruptions. 
While SHS enhances network efficiency, they do not inherently guarantee spanning resilience in the face of failures, which are common in complex networks. This limitation highlights the necessity to extend beyond the traditional focus on SHS nodes~\cite{lou2013mining, rezvani2015identifying, he2016joint} and emphasize not only strategic, inter-community connections but also resilience to node or link failures.

In this study, we aim to fill a gap in existing research by introducing the concept of \textit{robust spanners} (RS) -- nodes that demonstrate stability in spanning network communities even amidst component failures. RS nodes play a crucial role in preventing information silos and facilitating the dissemination of critical knowledge when communication pathways are compromised. 
This concept becomes especially significant in real-world networks, including the Internet, social interactions, power grids, and biological systems. These networks, susceptible to failures like node deletions, link disruptions, or cascading faults, highlight the limitations of relying solely on SHS. In online social networks, where quantifying robustness becomes intricate~\cite{casiraghi2020improving}, and in power grids prone to large blackouts triggered by cascading failures~\cite{zhang2015assessment,moreno2021network}, the necessity of RS is evident.
% RS nodes play a crucial role in preventing information silos and facilitating the dissemination of critical knowledge when communication pathways are compromised. 
% These RS nodes are essential in various real-world networks, including the Internet, social interactions, power grids, and biological systems, which are all prone to disruptions like node deletions and cascading faults.~\cite{zhang2015assessment,moreno2021network}.
Biological networks, with varying levels of robustness based on environmental noise, further emphasize the role of resilient connectors in overcoming intrinsic fragility~\cite{navlakha2014topological}. Thus, identifying RS nodes that are resilient to perturbations caused by failures, emerges as imperative across domains, ensuring the stability and functionality of complex networks under real-world challenges.

% We introduce a parallel approach to identifying robust spanner (RS) nodes within large-scale social networks. Our approach hinges on the utilization of an entropy-based scoring mechanism as a basis for calculating a node's \textit{robust spanning index} (RSI). RSI gauges a node's ability to form clusters that span across different network communities. Consequently, nodes classified as RS demonstrate a remarkable ability to connect nodes from diverse groups, all while maintaining resilience against potential node and link failures. The salient contributions of this work are summarized as follows.

In this paper, we introduce a novel entropy-based scoring mechanism for robust spanner detection. The proposed metric, termed \textit{Robust Spanning Index} (RSI), quantifies a node's ability to support clusters that span across different communities despite network evolution.
Given the inherent complexity of large and evolving, real-world social network systems~\cite{burchard2020scalable}, identifying spanners within these networks is both computationally demanding and time-consuming. To meet the computational and memory challenges associated with RS node identification in large-scale social networks, we propose a parallel algorithm and CUDA-based implementation. 
% To meet the computational and memory challenge associated with RSI identification in large and evolving real-world complex networks, especially social networks, we propose a parallel algorithm and SYCL-based implementation for identifying RS nodes in large-scale social networks. 
% We benchmark our approach against the only existing parallel algorithm designed for SHS detection, ESH~\cite{li2019distributed}. 

The salient contributions of this work are as follows.
%methodology for RS detection that employs an entropy-based scoring mechanism to calculate a node's \textit{robust spanning index} (RSI). This index evaluates a node's ability to support clusters that span across different network communities. Given that real-world complex systems, especially social networks, are large and continue to grow over time, identifying spanners within these networks is both computationally challenging and time-consuming. Nonetheless, to the best of our knowledge, ESH~\cite{li2019distributed} is the only existing parallel algorithm designed for SHS detection. Recognizing the importance of RS detection in real-world networks, we propose a parallel algorithm and SYCL-based implementation for identifying RS nodes in large-scale social networks. The salient contributions of this work are summarized as follows.
%%\vspace{-0.15in}
\begin{itemize}
    \item We introduce the concept of \textit{Robust Spanner (RS)} as nodes uniquely capable of maintaining connectivity among communities and bridging structural gaps despite topological disruption. We present an entropy-based scoring technique, termed \textit{Robust Spanning Index (RSI)}, to detect the vertices with high potential to act as RS nodes.
    
    \item For many-core architectures, we propose an efficient parallel algorithm that aims at RS identification in large networks optimized for both memory consumption and computational time. Additionally, we provide a CUDA-based implementation for Nvidia GPU architectures.
    % heterogeneous computing architectures.

    \item We also present a theoretical analysis of our parallel method, detailing its time complexity and optimization to highlight its efficiency and practical relevance.

    \item We utilize social network datasets to compare our method with state-of-the-art techniques. 
    % Our analysis indicates the capability of RSI to identify nodes 
    Our empirical analysis reveals RSI's efficacy in pinpointing nodes that demonstrate spanning capabilities comparable to state-of-the-art methods while exhibiting significantly superior robustness in the face of node and link failures.
    Our GPU implementation achieves 
    % a speedup of up to $2273 \times$ and 
    an average speedup of \textbf{$244 \times$} compared to existing spanner detection algorithms. 
    % These results underscore the applicability of our approach to large social networks undergoing perpetual evolution.
    
\end{itemize}

 The rest of the paper is organized as follows.
Section~\ref{sec:background} introduces preliminary concepts. Section~\ref{sec:proposed_approach} details our proposed technique for robust spanner scoring while 
Section~\ref{sec:parallel_algo} presents our parallel algorithm for robust spanner detection. Section~\ref{sec:optimization} discusses various optimizations to enhance the algorithm's efficiency.
Section~\ref{sec:implementation} describes a CUDA-based implementation of the proposed algorithm and Section~\ref{sec:results} presents the experimental analyses. 
Section~\ref{sec:refs} reviews existing research relevant to structural hole spanner detection and robustness metrics.
Finally, Section~\ref{sec:conclusion} concludes the paper.

%%\vspace{-0.1in}
\section{Preliminaries}
\label{sec:background}
%%\vspace{-0.05in}
% \subsection{Social Network Graph Model}

 Let $G = (V, E)$ be an undirected graph representing a social network, where $V$ is the set of nodes denoting the individuals and $E$ is the set of edges indicating the social ties. The degree of a node $u \in V$
%which measures the number of immediate connections, 
is denoted by $d(u)$, while the set of neighboring nodes is denoted by $N(u)$. The clustering tendency of a node $u$ is measured in terms of the maximum number of triangles $T(u)$ it can participate in, relative to its degree $d (u)$~\cite{saramaki2007generalizations}. It is calculated as:
% %\vspace{-0.05in}
\begin{equation}
\label{eq:clust}
    \mathbb{C}(u) = \frac{2 \cdot T(u)} {d(u) \cdot (d(u) - 1)}
\end{equation}

 A community in a complex network, denoted by a unique integral identifier (ID), represents subsets of nodes characterized by dense interconnections within the community and weaker connections to nodes outside the community. 
 In social networks, a community represents a group of individuals with common social interests.
 
 Our work focuses on non-overlapping communities where an individual can belong to only one community. 
 % Non-overlapping communities consist of nodes belonging to a single group while overlapping communities allow nodes to belong to multiple groups simultaneously. 
 Let $C(u)$ be the unique community ID of vertex $u$ and $C_{all} = \{C_1, \dots, C_k\}$ be the set of all communities in $G$. A vertex $u$ is called a community \textit{border vertex} if there exists at least one neighbor vertex $v \in N(u)$ such that $C(u) \neq C(v)$. This implies that $u$ lies on the boundary between two different communities, making it a point of interest for analyzing the structure and connectivity of networks, especially in the study of social networks. 
 Thus, a structural hole spanner is a community border vertex, that bridges gaps between otherwise disconnected communities. Its unique position allows it to control or facilitate the flow of information and resources across the network, giving it a potential advantage in terms of social capital and influence~\cite{burt1992}. We define \textit{robust spanner (RS)} as a community border vertex that spans communities through multiple communication paths, offering resilience amidst minor disruptions within the ego network. 

\section{Proposed Approach}
\label{sec:proposed_approach}

\begin{figure}[htp]
    \vspace{-0.2in}
    \centering \includegraphics[width=0.8\linewidth]{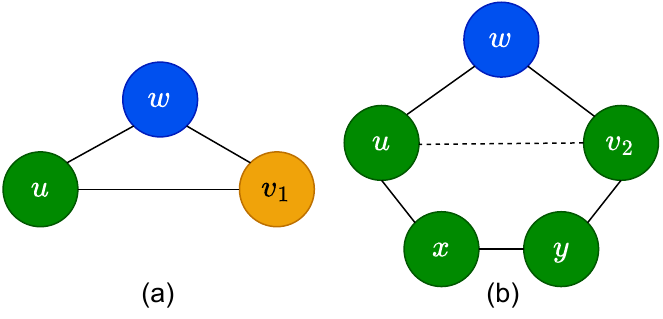}
    %%\vspace{-0.2in}
    \caption{Triad-finding for robust spanner detection: (a) \textit{type I}: closed triad $(u, w, v_1)$ and (b) \textit{type II}: open triad $(u, w, v_2)$}
    \label{fig:triad}
    \vspace{-0.15in}
\end{figure}

 % In this section, we focus on the essence of our proposed methodology, utilizing triads and entropy to identify Robust Spanners (RS) within social networks. This is premised on prior studies that show that the presence of triads, particularly \textit{feed forward loop} (FFL) motifs, serves as a measure of the closeness of nodes with their neighborhood and can be utilized to characterize network robustness due to the stability in local connections within clusters~\cite{heer2020maximising,roy2017role,roy2021influence,roy2020motifs}. 
 In this section, we present our proposed method, which leverages triads and entropy to identify Robust Spanners (RS) in social networks. Our approach builds on prior research demonstrating that the presence of triads, particularly \textit{feed-forward loop} (FFL) motifs, reflects the local cohesiveness of nodes within their neighborhoods. These motifs serve as indicators of network robustness by capturing the stability in local connections within clusters~\cite{heer2020maximising,roy2017role,roy2021influence,roy2020motifs}.
 This is evident from Fig. \ref{fig:triad}(a),  any single edge removal in the triad $(u,w,v_1)$ does not disrupt the information flow from node $u$ to $v_1$. We also discuss \textit{entropy} as a crucial measure of the diverse communities effectively connected by these identified robust spanners.

\subsection{Triad Formation}
\label{subsec:triadFormation}

 We consider two types of triangles during the robust spanner detection process. \textit{Type-I.} conventional triad comprising three nodes $u, v_1, w \in V$, where $u, w$, and $v_1$ belong to different network communities (illustrated using different colors in Fig. \ref{fig:triad}a); and \textit{Type-II.} an open triad consisting of $u, v_2, w \in V$ where $(u, v_2) \not \in E$, but still considered as a triangle because nodes $u, v_2$ belong to the same community as depicted by the green color and likely to be connected indirectly via other nodes (like $x, y$) in that community (Fig. \ref{fig:triad}b). In the context of this work, such multi-hop connections within a triad \textit{Type-II} are termed \textit{augmented edges}. While various other triad formations are possible, in the scope of our study, we recognize only \textit{Type-I} and \textit{Type-II} triads as valid triads.

% \begin{figure}[h!]
% % %\vspace{-0.3in}
% 	\centering
%         \subfloat[Triad 1]{	\includegraphics[width=0.25\linewidth]{figs/triad_formation1.png}
% 		\label{fig:triad1}
% 	}\hspace{1cm}
%         \subfloat[Triad 2]{%
%     \includegraphics[width=0.25\linewidth]{figs/triad_formation2.png}
% 		\label{fig:triad2}
% 	}
%  % %\vspace{-0.15in}
% 	\caption{Triad formations for robust spanner detection. The dashed line indicates a multi-hop connection between $u$ and $v$ belonging to the same community.}
% 	\label{fig:triadt}
% 	%\vspace{-0.2in}
% \end{figure}

%\subsection{An Weighted Graph Model:}

% \begin{figure}[h!]
% % %\vspace{-0.3in}
% 	\centering
%         \subfloat[Type-1]{	\includegraphics[width=0.25\linewidth]{figs/informationFlow1.png}
% 		\label{fig:infoFlow1}
% 	}\hspace{1cm}
%         \subfloat[Type-2]{%
%     \includegraphics[width=0.25\linewidth]{figs/informationFlow2.png}
% 		\label{fig:infoFlow2}
% 	}
%  % %\vspace{-0.15in}
% 	\caption{Information flow from $C(u)$ to other community in a triad. The directional arrows indicate the information flow in the triad.}
% 	\label{fig:infoFlow}
% 	%\vspace{-0.2in}
% \end{figure}

% \begin{equation}
%     W_v (u) = \mathbf{H} (L(u,v,w)) \times |L(u,v,w)|
% \end{equation}

% \begin{equation}
%     L(u, v, w) = \{ C(x) : x \in N(v), C(x) \neq C(u) \}
% \end{equation}

% \pagebreak 
%%\vspace{-0.05in}
\subsection{Robust Spanning Index}

 RS nodes can bridge \textit{different communities} within a network \textit{despite link failures}. Thus, a node $u \in V$ has a high robust spanning index (RSI) if its neighbor $v \in N(u)$.
 % \textcolor{red}{SD: illustrate every non-trivial concept and definition with examples}

\begin{enumerate}

    \item[a)] has neighbors ($N(v)$) connecting diverse communities, and

    \item[b)] participates in several local triangles with $v$, where the third node $x$ in the triangle co-occurs in the neighbor lists of $u$ and $v$, i.e., $\{ x:  x \in N(u) \cap x \in N(v) \} $.
    
\end{enumerate}

\subsubsection{Community diversity}\label{sec:diverse} Community diversity in a complex network refers to the extent of heterogeneity in the composition of communities, reflecting the presence of groups with different characteristics. The contribution of node $v$ towards the community diversity of a reference node $u$ is derived in terms of the diversity node $v$ offers to node $u$. 
% Specifically, we create a vector consisting of community IDs in the neighbor list of $v$ that do not possess the same community ID as node $u$, i.e., 
Specifically, we create a vector that counts how often each community, excluding that of node $u$, appears in the neighbor list of $v$, i.e., 
\begin{equation}\label{eq:cont}
%%\vspace{-0.05in}
    L(u, v) = \{ (C(x), freq(C(x))) : x \in N(v), C(x) \neq C(u) \}
\end{equation}

An element of set $L(u,v)$ is an ordered pair of the community ID of node $x$ and the frequency of the occurrence of its community ID, i.e., $freq(C(x))$. We use the entropy of $L(u,v)$ to determine the extent of diversity of communities attributed to node $u$ by the neighborhood of node $v$. 
% Given $p(c)$ representing the probability of observing community $c$ (proportional to its frequency $f = freq(c)$), the entropy of $L$ is:
Given $f_j = freq(C_j)$ representing the frequency of the 
 occurrence of community $C_j (1 \leq j \leq k)$ and $C(u) = C_i$, the probability of observing this community (denoted by 
 % p(f) = \frac{freq(c)}{\sum_{c^\prime freq(c^\prime)}}$
$p(f_j) = \frac{freq(C_j)}{\sum_{l = 1 \text{ to } k, l \neq i}freq(C_l)}$ 
 ) is proportional to its frequency and the entropy of $L(u,v)$ is:
%%\vspace{-0.02in}
\begin{equation}
\label{eq:entropy}
    \mathbf{H}(L(u,v)) = - \sum_{(C_j, f_j) \in L} p(f_j) \cdot \log(p(f_j))
\end{equation}
\subsubsection{Robustness to link failures} The robustness of a node $u \in V$ is its ability to participate in near-complete subnetwork (or triads) with $v, w \in N(v)$~\cite{heer2020maximising}. 
% This property is measured in terms of the weighted version of the clustering coefficient of node $u$ (see Equation \ref{eq:clust}) as: 
This property is quantifiable by the weighted version of the clustering coefficient of node $u$ (see Equation \ref{eq:clust}), referred to as RSI. It is represented as:
%%\vspace{-0.05in}
\begin{equation}\label{eq:RSI}
    \begin{aligned}
    \mathcal{R}(u) =\frac{1}{d(u) \cdot (d(u) - 1)} \sum_{w \in N(u), v \in N(w): \mathbbm{1}(u,w,v)}(\omega_v (u) \cdot \\ \omega_w (v) \cdot \omega_w (u)) ^{\frac{1}{3}}
    \end{aligned}
\end{equation}
\begin{equation}\label{eq:weight}
    \omega_v (u) = \mathbf{H} (L(u,v)) \cdot |L(u,v)|
\end{equation}
In Equation~\ref{eq:RSI}, $(\omega_v (u) \cdot \omega_w (v) \cdot \omega_w (u)) ^{\frac{1}{3}}$ is a weighted triad formed by nodes $u, v, w$ (See more details at  Section~\ref{subsec:informationFlow}.). The normalized weights $\omega$ and indicator function $\mathbbm{1}$ control the spanning potential of $u$. While $\omega_v(u)$ informs the model of the contribution of node $v$ in terms of community diversity towards the RSI of $u$, the indicator function $\mathbbm{1}$ 
% in Equation \ref{eq:RSI}
gauges whether $u, v$ are part of a valid triad that spans different communities, as follows: 
%%\vspace{-0.05in}
\begin{equation}
% %\vspace{-0.08in}
    \mathbbm{1}(u,w,v)=
    \begin{cases}
      1, & \text{if } (u,w,v) \text{ is a triad}, \\& C(v) \neq C(w), C(u) \neq C(w) \\
      0, & \text{otherwise}
    \end{cases}
% %\vspace{-0.05in}
\end{equation}

% \subsection{An Illustrative Example}

% Consider an example for the estimation of the robust spanning index (RSI) of node $u$ (in the toy network in Fig. \ref{fig:RSI}), which has neighbor nodes $w$ and $v$. While $v$ belongs to the same community as $u$, $w$ belongs to a different community (depicted by the different colors of the dotted circles). Further, the solid black color of a line connecting two nodes indicates the presence of a real link, whereas a dotted line shows that the nodes are connected by an undirected, multi-hop path by dint of being part of the same community. 

% As per Eq. \ref{eq:RSI}, the RSI of node $u$ depends on the community diversity offered by its neighbors $v, w$. Node $v$ has 2 neighbors ($x, y$), out of which $x$ belongs to the same (green) community as node $u$ and does not add to the diversity (given by Eq. \ref{eq:cont}). Thus, both $v$ and $w$ contribute a weight $\omega_v(u)$ and $\omega_w(u)$ of 1 by virtue of their respective neighbors $y$ and $z$ (see Eq. \ref{eq:weight}). The final step requires the normalization of the product of the weights of the valid robust spanning triad $(u, v, w)$ (equal to $1$), by $\frac{1}{d(u) \times (d(u) - 1)} = \frac{1}{2 \times (2 - 1)} = 0.5$, resulting in an RSI of $1 \times 0.5 = 0.5$ for node $u$.}

 % \begin{figure}[htp]
 %        %\vspace{-0.15in}
 %        \centering    \includegraphics[width=0.5\linewidth]{figs/robust_spanner.png}
 %        %\vspace{-0.1in}
 %        \caption{An illustration of a robust spanner.}
 %        \label{fig:robust_spanner}
 %        %\vspace{-0.1in}
 %    \end{figure}

%\vspace{-0.1in}
\section{Proposed Parallel Algorithm}
\label{sec:parallel_algo}
\begin{figure}[h!]
    \vspace{-0.3in}
    \centering \includegraphics[width=0.98\linewidth]{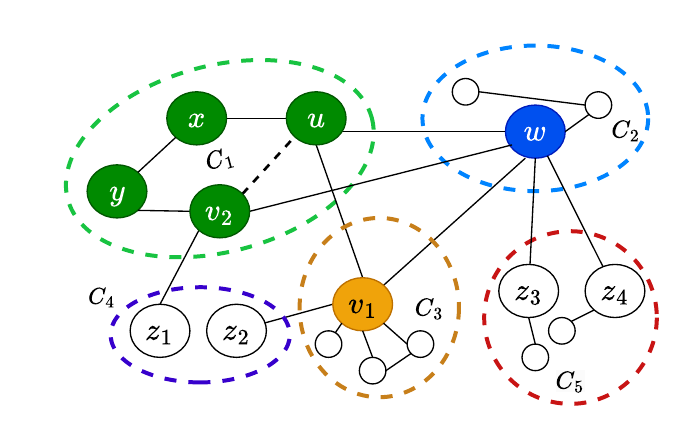}
    \vspace{-0.15in}
    \caption{An undirected social network $G(V,E)$ with a set of border vertices $V_b = \{u,v_1,v_2,w,z_1,z_2,z_3,z_4\}$.}
    \label{fig:RSI}
    % \vspace{-0.1in}
\end{figure}

% %\vspace{-0.05in}
% {\color{blue}
% \newpage
\noindent We devise a parallel algorithm (Algorithm~\ref{algo:Rscore}) to compute RSI for many core systems. Our approach hinges on the formation of triads comprising only community border vertices. In the example network shown in Figure~\ref{fig:RSI}, the triads of our interest are $(u,v_1,w)$ and $(u,v_2,w)$. Each vertex in any of these triads is a community-border vertex. In its initial step, Algorithm~\ref{algo:Rscore} reduces the computational domain by finding community border vertices, allowing us to focus solely on these vertices rather than the entire vertex-set $V$.

% \vspace{2pt}
 \textbf{Step 1: }It allocates all the vertices $V$ across available parallel threads. An asynchronous thread handling a particular vertex $u \in V$ traverses its neighbors $N(u)$ to find if there exist neighbors from a community distinct from $C(u)$. If such a neighbor is identified, vertex $u$ is classified as a community border vertex and subsequently added to the list of border vertices, $V_b$. At the end of this step, $V_b$ will contain $\{u,v_1,v_2,w,z_1,z_2,z_3,z_4\}$ for the network in Figure~\ref{fig:RSI}. 

 \textbf{Step 2:} If $(u,v,w)$ is a \textit{valid} triad (that bridges communities), the robustness of a vertex $u$ is calculated
using the contribution of vertex $v$ and $w$ to the diversity of the community, which influences the RSI of vertex $u$ (as detailed in Equation~\ref{eq:RSI}).
The diversity contribution of vertex $v$ towards RSI $\mathcal{R}(u)$ is denoted as $\omega_v(u)$. If a graph $G'$ is created using only the border nodes, $\omega_v(u)$  can be interpreted as the weight of a directed edge from a border vertex $v$ to another border vertex $u$ where the edge denotes the diversity contribution. Specifically, the second step of our algorithm constructs a graph $G'=(V_b, E_b)$ smaller than the original graph $G$, incorporating only the community border vertices $V_b$ and the non-zero diversity contribution (modeled as edges) towards each other.

\begin{figure}[htp]
    \vspace{-0.45in}
    \centering \includegraphics[width=0.85\linewidth]{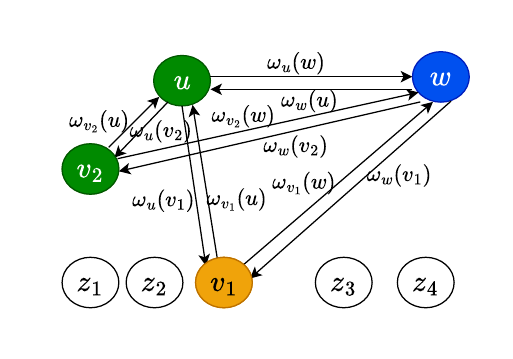}
    \vspace{-0.3in}
    \caption{Directed weighted network $G'(V_b,E_b)$.}
    \label{fig:graph_weighted}
    \vspace{-0.05in}
\end{figure}

For a vertex $u \in V_b$, this step identifies border vertices that are neighbors of $u$ or belong to the same community as $u$. Let $v \in V_b$ be such a neighbor of $u$. The algorithm then adds a directed edge $(v \rightarrow u)$ with weight $\omega_v(u)$ in $G'$. The computation of weight $\omega_v(u)$ depends on the entropy $H$ and it can be zero if all observed elements in $L(u,v)$ have a probability of one, or $v$ is linked only to one community. Entropy-based weighing measures a vertex $v$'s importance based on the number of communities it connects.

In general, the newly formed directed weighted network $G'$ is a subnetwork of $G$, and it helps the algorithm to operate efficiently during the subsequent graph traversal phase (i.e., Step 3). Step 2 converts the example network in Figure~\ref{fig:RSI} into the directed weighted network shown in Figure~\ref{fig:graph_weighted}. It contains only the border vertices and the edges with non-zero weights.

\begin{figure}[htp]
    \vspace{-0.1in}
    \centering \includegraphics[width=0.85\linewidth]{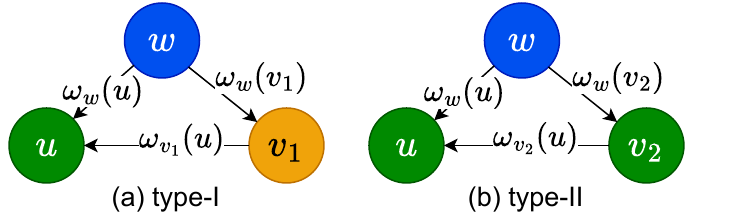}
    \vspace{-0.15in}
    \caption{Directed triad formation. Here $\omega_w(u)$ denotes the community diversity contribution of $w$ on $u$.}
    \label{fig:triad_weighted}
    \vspace{-0.08in}
\end{figure}

\vspace{2pt}
\noindent \textbf{Step 3: } The algorithm identifies the valid triads (see Section~\ref{subsec:triadFormation}) to compute the RSI $\mathcal{R}$.  Recall that Step 2 converts the undirected unweighted triads into directed weighted triads (Figure ~\ref{fig:triad_weighted}) to avoid redundant traversal to the neighbors that are not community border vertices.
Figure ~\ref{fig:triad_weighted}a and ~\ref{fig:triad_weighted}b illustrate examples of directed type-I triad $(u,v_1,w)$ and type-II triad $(u,v_2,w)$, respectively. 

In Step 3, for each directed edge $(v \rightarrow u)$, the algorithm identifies a common predecessor of $u$ and $v$ to detect a directed triad associated with $u$. Only triads that meet the criteria for type I or type II are selected, while other triads are dismissed. The $\mathcal{R}$ score (see Eq. \ref{eq:RSI}) for a vertex in $V_b$ is calculated by the parallel asynchronous threads.

While the existing triad detection approaches are based on triplet enumeration, linear algebra, and adjacency list intersection, we employ sorted adjacency lists for detection via sorted set intersection. Specifically, we perform intersections of sorted adjacency lists between nodes $u$ and $v$ using a merge-like method, incurring a cost of $O(d(u) + d(v))$ as presented in~\cite{fox2018fast}.

% In step 3, a vertex $u$ in $V_b$ is assigned to a parallel thread, which then visits the predecessor vertices $P_{G'}(u)$ of $u$. Let $v$ be a predecessor vertex of $u$. The algorithm identifies a common predecessor for both $u$ and $v$ to detect a directed triad associated with $u$. Only triads that meet the criteria for either type-I or type-II are selected, while other possible directed triads are dismissed. The $\mathcal{R}$-score (see Equation \ref{eq:RSI}) for a vertex in $V_b$ is calculated by the asynchronous threads in parallel.

\vspace{-0.1in}
\subsection{Information Flow in a Triad}
\label{subsec:informationFlow}
% \vspace{-0.05in}
The direction of the edges in a triad can be considered as the route for information inflow. Information from the community $C(w)$ can reach vertex $u$ of triad $(u,v,w)$ such that $C(u) \neq C(w), C(v) \neq C(w)$, via the direct path $w \rightarrow u$ or the indirect path $w \rightarrow v \rightarrow u$. Thus, removing an edge from the triad still ensures information reachability, preserving robustness.

% AK: Commented for now
% \textbf{Complexity Analysis:} Let $\mathcal{P}$ be the available parallel threads. Visiting the neighbors and finding if a neighbor belongs to a neighboring community takes $O(d_{avg})$ time in the average where $d_{avg}$ is the average degree of the input graph $G$. Therefore, finding all the border vertices in step 1 requires $O(\frac{|V|}{\mathcal{P}} \times d_{avg})$ time. 
% In step 2, the border vertices are distributed among the parallel threads (introduces a time factor of $\frac{|V_b|}{\mathcal{P}}$). Each thread then checks all the border vertices to find neighbors and it takes $O(|V_b|)$ time.
% Finding community diversity $L(u,v)$ in step 2 requires traversal to all the neighbors of $v$ in the original network. It takes $d_{avg}$ time. Computation of entropy $H$ and edge weight $\omega_v(u)$ takes $O(1)$ time. The amortized time for adding an edge $(u,v)$ in $E_b$ is also constant if stored in an adjacency matrix or list format. Finding the maximum edge weight using parallel reduction requires $O(\frac{|E_b|}{\mathcal{P}} \times log(|E_b|))$ time. The maximum theoretical value of $|E_b|$ is $2 \times \binom {|V_b|}2$. However, in practice, it is way smaller than this value as the number of edges crossing the community border is less.
% Therefore, the total time complexity of step 2 is $O(\frac{|V_b|}{\mathcal{P}} \times |V_b| \times d_{avg} + \frac{|E_b|}{\mathcal{P}} \times log(|E_b|)) = O(\frac{|V_b|^2}{\mathcal{P}} \times d_{avg} + \frac{|E_b|}{\mathcal{P}} \times log(|E_b|))$. (AK:not completed.)

\begin{algorithm}[h!]
% \SetAlgoLined
\DontPrintSemicolon
\caption{$R$-spanner ($G, C$)}
\label{algo:Rscore}

\tcc{Step 1: Filter border vertices}
Initialize an empty list $V_b$\;
\For{each vertex $u\in V$ in parallel}{
    \For{each neighbor $x \in N(u)$}{
        \If{$C(u) \neq C(x)$ }{
            Add $u$ to $V_b$\;
            break\;
        }
    }
}
\tcc{Step 2: Weight Computation}
Initialize an empty edge set $E_b$\;
\For{each vertex $u\in V_b$ in parallel}{
    \For{each vertex $v \in V_b$}{
        \If{$C(u) = C(v)$ or $v \in N(u)$}{
            Find $L(u,v)$ \label{code:find_L}\;
            \If{$|L(u,v)| > 1$}{
                Compute $H(L(u,v))$ \label{code:compute_H}\;
                $\omega_v(u) \gets H(L(u,v)) \times |L(u,v)|$\;
                Add a directed edge $(v \rightarrow u)$ with weight $\omega_v(u)$ to $E_b$\;
            }
        }
    }
}
Create a graph $G'= (V_b,E_b)$\;
Find max edge weight $\omega_{max}$ in $E_b$ using parallel reduction\;

\tcc{Step 3: RSI $(\mathcal{R})$ computation}
Initialize an array $\mathcal{R}$ of size $|V|$ with value $0$\;
% \For{each vertex $u \in V_b$ in parallel}{
%     $\mathcal{R}(u) \gets 0$\;
%     \For{each predecessor $v \in P_{G'}(u)$}{
%         \For{each predecessor $w \in P_{G'}(v)$}{
%             \If{$w \in P_{G'}(u)$ and $C(w) \neq C(v)$}{
%                 $\mathcal{R}(u) \gets \mathcal{R}(u) + (\omega_v (u) \times \omega_w (v) \times \omega_w (u)) ^{\frac{1}{3}}/ \omega_{max}$\;
%             }
%         }        
%     }
%     $\mathcal{R}(u) \gets \frac{\mathcal{R}(u)}{(d(u) \times (d(u) - 1))}$\;
% }
\For{each directed edge $(v \rightarrow u) \in E_b$ in parallel}{
    \For{each $w$ being a common predecessor of both $u$ and $v$}{
        \If{$C(w) \neq C(v)$}{
            $\mathcal{R}(u) \gets \mathcal{R}(u) + (\omega_v (u) \times \omega_w (v) \times \omega_w (u)) ^{\frac{1}{3}}/ \omega_{max}$\;
        }      
    }
}
\For{each vertex $u \in V_b$ in parallel}{\label{code:R_div}
    $\mathcal{R}(u) \gets \frac{\mathcal{R}(u)}{(d(u) \times (d(u) - 1))}$\;
}

\tcc{(Optional) Step 4: Find  top-$K$ vertices with highest $\mathcal{R}$ values}

\end{algorithm}

% \begin{algorithm}
% \SetAlgoLined
% \DontPrintSemicolon
% \SetKwFunction{computeR}{computeR}
% \SetKwProg{Pn}{}{:}{\KwRet $R$}
% \Pn{\computeR{$G_B (V_b, E_b), C, R$}} {

% \caption{ComputeR function}
% \label{algo:computeR}

% \For{each vertex $u \in V_b$ in parallel}{
%     $R(u) \gets 0$\;
%     \For{each predecessor $v \in P_{G_{B}}(u)$}{
%         \For{each predecessor $w \in P_{G_{B}}(v)$}{
%             \If{$w \notin P_{G_{B}}(u)$ or $C(w) = C(v)$}{
%                 continue\;
%             }
%             % \If{$C(u) \neq C(v)$}{
%             %     \tcc{ Compute for triad type 1}
%             % }
%             % \Else{
%             %     \tcc{ Compute for triad type 2}
%             % }
%             \Else{
%                 $R(u) \gets R(u) + (\omega_v (u) \times \omega_w (v) \times \omega_w (u)) ^{\frac{1}{3}}$\;
%             }
%         }        
%     }
%     $R(u) \gets R(u) / (d(u) \times (d(u) - 1))$\;
% }
% }
% \end{algorithm}

%\vspace{-0.05in}
\section{Proposed Optimizations}
\label{sec:optimization}

 Let us discuss how the parallel RS approach achieves performance optimization in terms of memory and computation.
% {\color{blue}
\vspace{-0.05in}
\subsection{Memory Optimization}
Our implementation targets GPUs, which typically have much less memory than system RAM.

\begin{lemma}
\label{lemma:1}
\vspace{-0.05in}
    For any two directed edges $(v \rightarrow u_1)$ and $(v \rightarrow u_2)$ 
    , if $u_1$ and $u_2$ belong to the same community then the edge weights $\omega_v (u_1)$ and $\omega_v (u_2)$ in $G'=(V_b, E_b)$ will be the same.
\end{lemma}
\begin{proof}
    % two possible cases: 1. v belongs to a different community
    % 2. v belongs to the same community
    As per Equation~\ref{eq:weight}, edge weight of $(v \rightarrow u_1)$ is 
    \vspace{-0.05in}
    \begin{multline}
    \vspace{-0.15in}
        \omega_v (u_1) = \mathbf{H} (L_1) \times |L_1|,\text{ where }
        \\L_1 = L(u_1, v) = \{ (C(x), freq(C(x))) : x \in N(v),\\ C(x) \neq C(u_1) \}.
    \vspace{-0.15in}
    \end{multline}
    Similarly, the edge weight of $(v \rightarrow u_2)$ is 
    \vspace{-0.05in}
    \begin{multline}
    \label{eq:omega_v (u_2)}
    \vspace{-0.15in}
        \omega_v (u_2) = \mathbf{H} (L_2) \times |L_2|,\text{ where }
        \\L_2 = L(u_2, v) = \{ (C(x), freq(C(x))) : x \in N(v),\\ C(x) \neq C(u_2) \}. 
    \end{multline}
    Nodes $u_1$ and $u_2$ belong to the same community $C(u_1) = C(u_2)$. Substituting $C(u_2)$ by $C(u_1)$ in Equation~\ref{eq:omega_v (u_2)}, we get $L_2 
    =\{ (C(x), freq(C(x))) : x \in N(v), C(x) \neq C(u_1)\} 
    = L_1$. Or, $\omega_v (u_2) 
    = \mathbf{H} (L_2) \times |L_2| 
    = \mathbf{H} (L_1) \times |L_1| = \omega_v (u_1)$. 
%\vspace{-0.05in}
\end{proof}

\begin{lemma}
\vspace{-0.05in}
\label{lemma:2}
    The weight of any augmented directed edge starting from $v$ in any Type II triad can be computed independently of the other endpoint of the edge.
\end{lemma}
\begin{proof}
    Let there be $y$ number of Type II triads consisting of $v$ in $G'$. Let the augmented edges in those triads be respectively $(v \rightarrow u_1), \dots, (v \rightarrow u_y)$. As per Lemma~\ref{lemma:1}, the edge weights are the same, i.e., $\omega_v (u_1) = \dots = \omega_v (u_y)$. As per Equation~\ref{eq:weight}, 
    \begin{multline}
    \vspace{-0.15in}
        \omega_v (u_i) = \mathbf{H} (L) \times |L|,\text{ where } i = 1,\dots,y,\text{ and }
        \\L = L(u_i, v) = \{ (C(x), freq(C(x))) : x \in N(v),\\ C(x) \neq C(u_i) \}.
    \vspace{-0.15in}
    \end{multline}
    By definition of augmented edge in triad Type II, $v$ and $u_i$ belong to the same community. Hence, $C(u_i) = C(v)$ and $L(u_i, v) = \{ (C(x), freq(C(x))) : x \in N(v), C(x) \neq C(v)\} = L(v)$, i.e., the computation of $L$ or $\omega_v (u_i)$ is independent of $u_i$.
\end{proof}
\vspace{-0.05in}
% \textcolor{red}{Illustrate Lemmas and Theorem(s) w.r.t. Figure 1, else via schematic diagrams.}
Lemma~\ref{lemma:1} establishes that the edge weight $\omega_v(u)$ is solely determined by the community $C(u)$ and neighborhood of $v$. As shown in Figure~\ref{fig:graph_weighted}, since $C(u) = C(v_2) = C_1$, the edge weights $\omega_w(u)$ and $\omega_w(v_2)$ are identical.
Lemma~\ref{lemma:2} clarifies that for an augmented edge, the weight $\omega_v(u)$ is based on either $C(v)$ or $C(u)$, given they are equivalent. Combining insights from Lemmas~\ref{lemma:1} and~\ref{lemma:2}, we infer that if multiple edges or augmented edges are present from a border vertex $v$ to a community $C_i$ (or the vertex's own community $C(v)$) in $G'$, a single variable suffices to store these edge weights. Henceforth, we will use $\omega_v(C_i)$ to denote the edge weight of any directed edge from a border vertex $v$ to any other vertex belonging to the community $C_i$.

\vspace{-0.075in}
\begin{lemma}
\label{lemma:3}
    Let there be $\eta_v$ number of communities (including the community of $v$) connected to a border vertex $v$ in $G'$. Then any directed edge $(v \rightarrow u)$ may have at most $\eta_v$ distinct edge weights (with the exact number depending on the community of $u$).
\end{lemma}
\begin{proof}
    The weight of an augmented edge $(v \rightarrow u)$ depends on the community of $v$ and not on $u$ (Lemma~\ref{lemma:2}). The weight of any other edge $(v \rightarrow u)$ in a triad is independent of all properties of $u$ besides its community ID, i.e., $C(u)$ (Equation~\ref{eq:cont}). 
    % Therefore, an edge starting from $v$ can have an edge weight $\omega_v (u) = \mathbf{H} (L) \times |L|$ where $L = L(u, v) = \{ (C(x), freq(C(x))) : x \in N(v), C(x) \neq C(u) \}$. 
    As $v$ is connected to $\eta_v$ number of communities (including that of $v$) only, $C(u)$ can have $\eta_v$ distinct values, which in turn can create $\eta_v$ distinct edge weights using Equation~\ref{eq:cont}, and~\ref{eq:weight}.
\end{proof}

\vspace{-0.2in}
\subsection{Computational Optimization}
\label{subsubsec:computational_optimization}
An edge weight $\omega_v(C_i)$ depends on  $p(freq(C_i))$, i.e., the probability of observing the community $C_i$ in the neighborhood of $v$. In this section, we identify computations that are shared across multiple edge weight calculations.
Later in our implementation, we store the values of these shared calculations to avoid redundant computation.

If there exists $k$ communities, an edge from a border vertex $v$ to another vertex in $G'$ will have an edge weight among $\{\omega_v(C_1), \dots, \omega_v(C_k)\}$. An edge weight $\omega_v(C_i) = H(L_i) \times |L_i|$, where $H(L_i)$ and $|L_i|$ follow Equations~\ref{eq:cont} and~\ref{eq:entropy}. 
\begin{multline}
\label{eq:opt_H}
H(L_i) = - \sum_{(C,f) \in L_i}p(f) \times \log(p(f)) 
\\= -\sum_{j = 1 \text{ to } k, j \neq i} p(freq(C_j)) \times \log(p(freq(C_j)))
\end{multline}
Here, the probability $p$ can be written as follows:
\begin{multline}
\label{eq:p}
p(freq(C_j)) = \frac{freq(C_j)}{\sum_{l = 1 \text{ to } k, l \neq i}freq(C_l)}
= \frac{f_j}{Y_i} 
\\\text{ [Let } f_j = freq(C_j) \text{ and } Y_i = \sum_{l = 1 \text{ to } k, l \neq i}f_l 
\\\text{ or, } Y_i = \sum_{l = 1}^k f_l - f_i
= \mathcal{T} - f_i \text{, where } \mathcal{T} = \sum_{l = 1}^k f_l\text{]}
\end{multline}
Using Equations \ref{eq:opt_H} and~\ref{eq:p}, we get 
\begin{multline}
\label{eq:H_optimal}
H(L_i) = -\frac{1}{Y_i}\sum_{j = 1 \text{ to } k, j \neq i} f_j \times \log(\frac{f_j}{Y_i})
\\= -\frac{1}{Y_i}(\sum_{j = 1}^k f_j \times \log(\frac{f_j}{Y_i}) - f_i \times \log(\frac{f_i}{Y_i}))
\\= -\frac{1}{Y_i}(\sum_{j = 1}^k f_j \times \log(f_j) - \sum_{j = 1}^k f_j \times \log(Y_i) - f_i \times \log(\frac{f_i}{Y_i}) )
\\= -\frac{1}{\mathcal{T} - f_i}(\mathcal{X} -  \mathcal{T} \times \log(\mathcal{T} - f_i) - f_i \times \log(\frac{f_i}{\mathcal{T} - f_i}) )
\end{multline}
From Equation~\ref{eq:H_optimal}, it is evident that both $\sum_{j = 1}^k f_j \times \log(f_j)$ (Let $\mathcal{X}$) and $\mathcal{T}$ ($= \sum_{j = 1}^k f_j$) do not depend on $i$. $H(L_i)$ becomes a function of $\mathcal{X}, \mathcal{T}$, and $f_i$. Hence, by preserving the values of $\mathcal{X}$, and $\mathcal{T}$ we can circumvent redundant computations when determining $\omega_v(C_i)$. 
% }

%\vspace{-0.1in}
\section{Parallel Implementation}
\label{sec:implementation}
% \begin{itemize}
%     \item Graph adjacency list stored in CSR format (requires  less memory, easy to work on 1D flattened array in GPU)
%     \item global vertex id and border vertex id map using a simple integer array. (There is no in-built map datatype in SYCL) 
%     \item fetch\_add based filter (try to improve using prefix sum)
%     \item SYCL 2020 specification-based reduction operation to find max weight in step 3B.
%     \item Store $W[v][v]$ as edge weight from any $u \in V_b$ such that $C(u) = C(v)$. (Reduces total memory requirement)    
% \end{itemize}

\subsection{Data Structure}
 The input graph's adjacency list is stored in Compressed Sparse Row (CSR) format, where a row pointer array indicates the start of the neighbor list for each vertex, while a 1D flattened neighbor list array holds all neighboring vertices in row-major order. We utilize an integer array $bv$ of length $|V_b|$ to maintain the global vertex  ID of each border vertex, thereby mapping a local ID of a border vertex to its global ID. Conversely, another array $bv\_local$, sized $|V|$, is used to map global vertex IDs to the local IDs of border vertices. For vertices that aren't border vertices, -1 is stored in the array.

Generalization of Lemma~\ref{lemma:3} implies that an edge (in $G'$) originating from a border vertex $v$ to another border vertex can have at most $k$ different edge weights when $k$ is the total number of communities. Therefore, it requires $|V_b| \cdot k$ space to store all the edge weights. Also, a vertex that connects to larger communities plays a key role in the diffusion of information across the graph~\cite{li2019distributed}. Hence, the spanner detection techniques take advantage of this property and directly try to search the spanners in user-provided large communities only~\cite{lou2013mining}. 

In our implementation, the community IDs are considered user input. Given that there are $k$ communities to search, compute and store the edge weights $\omega_v(C_i)$ ($1 \leq i \leq k$), we leverage a 2D data structure (named vertex-community interaction matrix and denoted by $\myDS$)  of size $|V_b| \cdot k$, whose $(j, i)^{th}$ element is used to compute and store the edge weight $\omega_v(C_i)$ for the $j^{th}$ border vertex in $bv$. 
% It is noteworthy that only the last column is dedicated to storing the total neighbors $\mathcal{T} = \sum_{l = 1}^kf_l$ of the border vertices belonging to the $k$ considered communities. 
For ease of edge weight computation, we store the total number of neighbors $\mathcal{T} = \sum_{l = 1}^k f_l$ belonging to the $k$ considered communities in a separate array of size $|V_b|$.

%\vspace{-0.1in}
\subsection{CUDA Kernels}
\begin{figure}[htp]
    \vspace{-0.1in}
    \centering \includegraphics[width=0.88\linewidth]{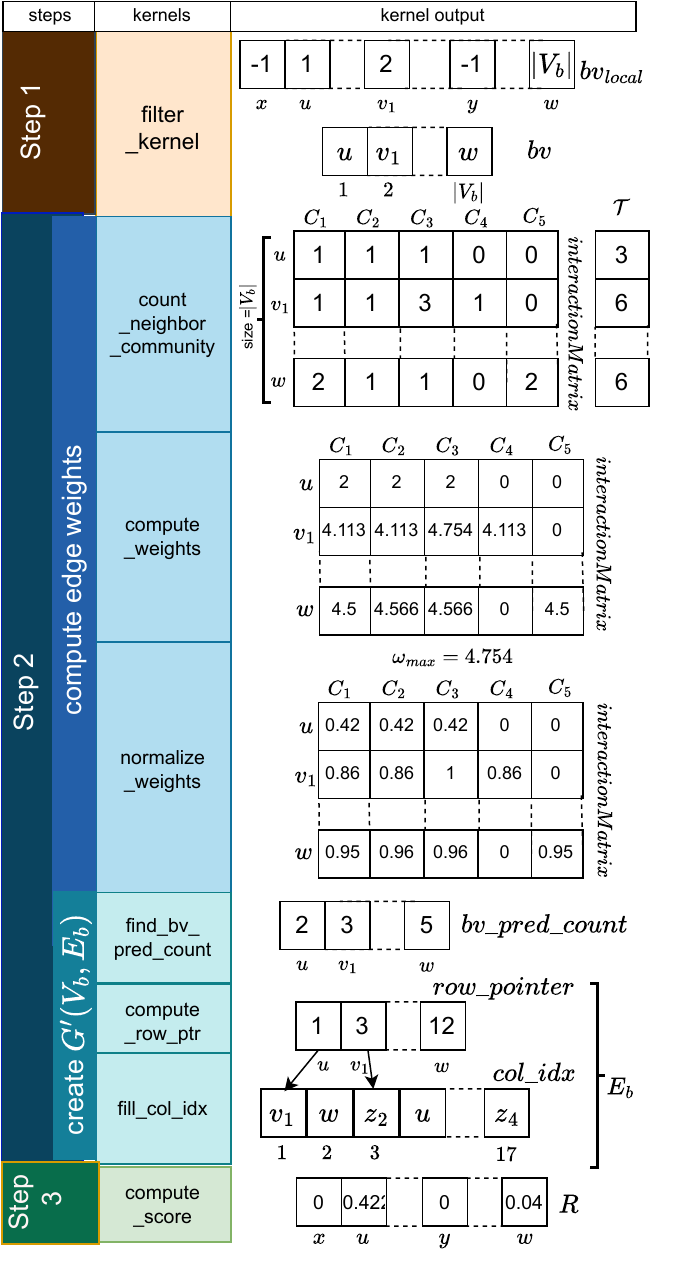}
    \vspace{-0.15in}
    \caption{Kernel workflow. 
    % \textcolor{red}{SD: displaying this Fig horizontally spanning both columns will save space}
    }
    \label{fig:kernels}
    %\vspace{-0.1in}
\end{figure}

 For efficient GPU implementation, we divide Algorithm~\ref{algo:Rscore} into different CUDA kernels.\\
\noindent \textbf{filter\_kernel:} It executes step 1 of the algorithm, populating the $bv$ and $bv\_local$. 
Populating the $bv$ array without using atomic operations requires array compaction, which involves marking the border vertices and efficiently storing them in the compacted array $bv$. To optimize this process, we employ \texttt{\_\_ballot\_sync()} for warp-wide voting, generating 32-bit masks that indicate the identified border vertices. Each thread then uses the \texttt{\_\_popc()} function to compute its offset, which represents the number of valid elements preceding it within the warp. Finally, threads use these offsets to store the border vertices efficiently in the $bv$ array.
% It uses SYCL's lightweight atomic operation $fetch\_add$ to count border vertices. 
Fig.~\ref{fig:kernels} illustrates various kernel operations applied to the graph from Fig.~\ref{fig:RSI}. The \textit{filter\_kernel} detects the border vertices $u,v_1,v_2,z_1,z_2,z_3,z_4,w$ and saves them in $bv$. Also, a unique local ID associated with the indices of $bv$ is attributed to each border vertex and recorded in $bv\_local$.

Step 2 of Algorithm~\ref{algo:Rscore} is broken down into two sub-steps: 1) compute edge weights, 2) create $G'$. Based on Lemma~\ref{lemma:1} and ~\ref{lemma:2}, edge weight $\omega_v(C_i)$ in $G'$ is solely determined by $v$'s neighbors and community data in the original graph $G$. This allows us to calculate edge weights in $G'$ even before identifying the actual directed edges. We use the following three kernels to compute them.

\noindent \textbf{count\_neighbor\_community:} This kernel iterates through the neighbors of each border vertex, counting the number of neighbors belonging to each community and storing this information in $\myDS$.
Since each thread writes to a separate row of $\myDS$, no atomic operations are needed. After execution, $\myDS(j, i)$ holds the count of neighbors from community $C_i$ for the $j^{\text{th}}$ border vertex. The total count across all $k$ communities is stored in $\mathcal{T}$.

\noindent \textbf{compute\_weights:} This kernel leverages the computational optimization proposed in Section~\ref{subsubsec:computational_optimization} to efficiently compute edge weights, as outlined in Algorithm~\ref{algo:compute_weights}.

\begin{algorithm}
% \SetAlgoLined
\DontPrintSemicolon
\caption{Compute weights}
\label{algo:compute_weights}

Initialize $\mathcal{X}, L_{all}$ to $0$\;
\For{ $j = 1$ to $|V_b|$ in parallel}{
    \For{each community $C_i$}{
        $f_i \gets \myDS(j,i)$\;
        \If{$f_i > 0$}{
            $\mathcal{X} \gets \mathcal{X} + f_j \cdot \log(f_j)$\;
            % $\mathcal{X}_2 \gets \mathcal{X}_2 + f_j$\;
            $L_{all} \gets L_{all} + 1$\;
        }
    }
    $\mathcal{T} \gets \myDS(j,(k+1))$\;
    $L \gets L_{all} - 1$\;
    \For{each community $C_i$}{
        $f_i \gets \myDS(j,i)$\;
        $Y \gets \mathcal{T} - f_i$\;
        $H \gets -\frac{1}{Y}(\mathcal{X} - \mathcal{T} \cdot \log(Y) - f_i \cdot \log(\frac{f_i}{Y}) )$\;
        $\omega \gets H \cdot L$\;
        $\myDS(j,i) \gets \omega$\;
    }
}
\end{algorithm}

% \noindent \textbf{find\_max\_weight:} It uses SYCL 2020 specification-based parallel reduction operation to find the maximum weight from the $\myDS$ and stores in $\omega_{max}$.\\
\noindent \textbf{normalize\_weights:} 
This kernel parallelizes the normalization of weights in $\myDS$ by dividing each element by the maximum value $\omega_{\text{max}}$. We use \texttt{thrust::max\_element} to compute $\omega_{\text{max}}$. The division is parallelized with \texttt{thrust::transform}.

In Fig.~\ref{fig:RSI}, vertex $u$ has neighbors $x, v_1,$ and $w$ in communities $C_1, C_3,$ and $C_2$, respectively. Fig.~\ref{fig:kernels} depicts the process by which $\myDS$ captures this neighborhood information and gradually calculates the normalized weights. For vertex $u$,  $\mathcal{T} = 3$ and $\mathcal{X} = 3 \cdot (1 \cdot \log(1)) = 0$. $H(L_1) = \frac{1}{3 - 1} \cdot (0 - 3 \cdot \log(3-1) - 1 \cdot \log (\frac{1}{3-1})) = 1$. Hence, the weight before normalization $\omega_u'(C_1) = H(L_1) \cdot |L| = 1 \cdot |L_{all} - 1| = 2$. In our example $\omega_{max} = 4.754$ and therefore the normalized weight will be $\omega_u(C_1) = 2/4.754 = 0.42$.

The next sub-step in Step 2 constructs $G'(V_b, E_b)$, where $V_b$ consists only of border vertices. In $G'$, a directed edge $(v \rightarrow u) \in E_b$ exists only if both $v$ and $u$ are border vertices. For any pair of border vertices $u, v$ within the same community $C_j$, the augmented edges $(u \rightarrow v)$ and $(v \rightarrow u)$ can be inferred during triad-II computation by simply checking their community membership. Hence, to reduce space complexity, we include in $E_b$ only those edges $(v \rightarrow u)$ that satisfy: 1) $(u, v) \in E$, 2) both $u$ and $v$ are in $V_b$, and 3) $C(u) \neq C(v)$.
\\
\noindent \textbf{find\_bv\_pred\_count:}
For each border vertex $v \in V_b$, this kernel traverses its neighbor list in the original graph $G$ to identify potential predecessors in $G'$. A neighbor of $v$ in $G$ qualifies as a predecessor in $G'$ if it is also a border vertex. The total number of such predecessors for each $v \in V_b$ is stored in an auxiliary array \textit{bv\_pred\_count} of size $|V_b|$.\\
\noindent \textbf{compute\_row\_ptr:} It employs \texttt{thrust::exclusive\_scan} to generate the row pointer of $G'$ CSR. The $i^{th}$ element points to the starting position of the neighbor list ($col\_idx$) for the $i^{th}$ border vertex in $G'$ (see Fig.~\ref{fig:kernels}).\\
\textbf{fill\_col\_idx:} Using the row pointer from the previous kernel, we populate the neighbor list of $G'$ in parallel by distributing border vertices across threads. Since Algorithm~\ref{algo:Rscore} step 3 requires only predecessors, we store only predecessor entries in the neighbor list. Given that the number of predecessors per vertex can vary significantly, grid-stride loops are employed for better load balancing.

Finally, the following kernel implements Step 3 of Algorithm~\ref{algo:Rscore}.\\
\noindent \textbf{compute\_score:}
This kernel uses the directed topology of $G'$ and community information to identify directed triads. Triad detection employs a merge-based method to efficiently find common neighbors of each directed edge’s endpoints~\cite{fox2018fast}. It then computes the $R$-scores using edge weights from $\myDS$. Grid-stride loops are used to balance the irregular workloads throughout the kernel. Despite incorporating a fast triad detection technique, this remains the most compute-intensive kernel.

% Directed triad detection uses a merge-like method to find the common neighbors of each directed edge~\cite{fox2018fast}.

% Changing(AK)
% Fig.~\ref{fig:kernels} illustrates various kernel operations applied to the graph from Fig.~\ref{fig:RSI}. 
% In Fig.~\ref{fig:RSI}, vertex $u$ has neighbors $x, v_1,$ and $w$ in communities $C_1, C_3,$ and $C_2$, respectively. $\myDS$ captures this neighborhood information and computes the weights for the edges originating from $u$ (see Fig.~\ref{fig:kernels}).\\
% In Fig.~\ref{fig:kernels}, for vertex $u$,  $\mathcal{T} = 3$ and $\mathcal{X} = 3 \times (1 \times log(1)) = 0$. 
% $H(L_1) = \frac{1}{3 - 1} \times (0 - 3 \times log(3-1) - 1 \times log (\frac{1}{3-1})) = 1$. 
% Hence, $\omega_u(C_1) = H(L_1) \times |L| = 1 \times |L_{all} - 1| = 2$. 
% $G'$ captures the edges among the border vertices only and helps to find the triads related to $u$ efficiently.
In Fig.~\ref{fig:kernels}, the last step uses $G'$ and detects triads $(u,v_1,w)$, $(u,w,v_1)$ of type-I and $(u,v_2,w)$ of type-II. Hence, the RSI of $u$ is computed as $\frac{1}{3 \cdot 2} \cdot ((0.95 \cdot 0.96 \cdot 0.86)^\frac{1}{3} + (0.95 \cdot 0.95 \cdot 0.42)^\frac{1}{3} + (0.86 \cdot 0.86 \cdot 0.95)^\frac{1}{3}) = 0.422$.

%\vspace{-0.1in}
\subsection{More Analysis on Proposed Method}
In Algorithm~\ref{algo:Rscore}, Step 1 examines each vertex's neighbor list to identify border vertices, requiring $O(\frac {|V| \cdot d_{max}}{\rho})$ time, where $\rho$ denotes the available parallel processors and $d_{max}$ is the maximum degree of graph $G$. Algorithm~\ref{algo:compute_weights} optimizes computing Step 2, necessitating $O(\frac{|V_b| \cdot k}{\rho})$ time, with $|V_b|$ representing the number of border vertices and $k$ the total communities. Parallel reduction for maximum weight finding and normalization contributes $O(\frac{|V_b| \cdot k}{\rho} + \log(\rho))$ time. For triad detection and RSI computation in Step 3, a merge-like technique~\cite{fox2018fast} is utilized, performing $O(d(u)+d(v))$ work for each edge $(v \rightarrow u) \in E_b$ and requiring up to $O(\frac{|E_b| \cdot (2 \cdot d'_{max})}{\rho} + \frac{|V_b|}{\rho})$ time, where the latter term comes from Algorithm~\ref{algo:Rscore} Line~\ref{code:R_div} and $d'_{max}$ indicates maximum degree of $G'$. 
Thus, the total time complexity is $O(\frac {|V| \cdot d_{max}}{\rho} + \frac{|V_b| \cdot k}{\rho} + \frac{|V_b| \cdot k}{\rho} + \log(\rho) + \frac{|E_b| \cdot (2 \cdot d'_{max})}{\rho} + \frac{|V_b|}{\rho}) 
=
O(\frac {|V| \cdot d_{max}}{\rho} + \frac{|V_b| \cdot k}{\rho} + \log(\rho) + \frac{|E_b| \cdot d'_{max}}{\rho})$. 
Given that $G'$ is generally sparse and the relationship $O(E_b) \equiv O(V_b)$ holds, along with $k \leq d'_{max}$, the overall time complexity simplifies to  $O(\frac {|V| \cdot d_{max}}{\rho} + \log(\rho) + \frac{|V_b| \cdot d'_{max}}{\rho})$. 

The algorithmic complexity depends on the number of border vertices ($|V_b|$). Next, we discuss the likelihood of a vertex being a border vertex for a given inter-community link probability.

\begin{lemma}
\label{lemma:4}
%\vspace{-0.05in}
    In a network with the sizes smallest and largest communities given by $n_S$ and $n_L$, respectively, and the probability of the existence of inter-community link be $p$, the expected probability of a node being a border vertex $p_{border}$ lies between $1 - e^{- (|V| - n_L) \cdot p}$ and $1 - e^{- (|V| - n_S) \cdot p}$.  
\end{lemma}
\begin{proof}
%\vspace{-0.1in}
    Consider a network $G = (V, E)$ with a probability of inter-community link existence $p$. Two nodes $u \in V$ and $v \in V$ belong to the smallest community $c_S$ (with $n_S$ nodes) and largest community $c_L$ ($n_L$ nodes), respectively.
    % (Fig. \ref{fig:border}).  

    %  \begin{figure}[htp]
    %     % %\vspace{-0.15in}
    %     \centering \includegraphics[width=0.6\linewidth]{figs/border.png}
    %     %\vspace{-0.15in}
    %     \caption{A network with 3 communities, where the largest community $c_L$ (green dashed ellipse) and the smallest community $c_S$(red dashed ellipse) have $n_L$ and $n_S$ nodes respectively. The dotted lines are inter-community links.}
    %     \label{fig:border}
    %     % %\vspace{-0.2in}
    % \end{figure}

    For any node in $G$, let the total possible inter-community links represent an interval and the number of existing inter-community links denote \textit{independent} successes within the interval. Then, the expected number of inter-community links $X$ in $G$ follows the Poisson distribution. For node $u$ 
    % in Fig. \ref{fig:border}
    , the expected number of inter-community links is given by $\lambda_S = p \cdot (|V| - n_S)$. Since, $P(x; \lambda) = \frac{e^{-\lambda} \cdot \lambda^x}{x!}$, the probability of $u$ being a border vertex (or having at least one neighbor belonging to a different community) is given by: 

%\vspace{-0.1in}
    \begin{equation}
        P(X \geq 1) = 1 - P(X = 0) = 1 - e^{-\lambda_S} = 1 - e^{- (|V| - n_S) \cdot p}
    \end{equation}

    \noindent Similarly, the probability of $v$ 
    % in Fig. \ref{fig:border} 
    being a border vertex: 

%\vspace{-0.15in}
    \begin{equation}
        P(X \geq 1) = 1 - P(X = 0) = 1 - e^{-\lambda_L} = 1 - e^{- (|V| - n_L) \cdot p}
    \end{equation}

% %\vspace{-0.05in}
    \noindent Nodes $u$ and $v$ belong to the smallest and largest community with the maximum number of neighbors $|V| - n_S$ and $|V| - n_L$, respectively. Since the number of inter-community neighbors of all nodes must lie in the interval $(|V| - n_L, |V| - n_S)$, the expected probability ($p_{border}$) of a node to be a border vertex  will lie between $1 - e^{- (|V| - n_L) \cdot p}$ and $1 - e^{- (|V| - n_S) \cdot p}$, i.e.,  
    \begin{equation}
        1 - e^{- (|V| - n_L) \cdot p} \leq p_{border} \leq 1 - e^{- (|V| - n_S) \cdot p}
        %\vspace{-0.05in}
    \end{equation}
% %\vspace{-0.2in}
\end{proof}

\begin{table}[htbp!]
% %\vspace{-0.15in}
\caption{Social Networks Dataset (Small)~\cite{nr}}
%\vspace{-0.05in}
\label{table:graph_small}
\centering
% \resizebox{\textwidth}{!}{%
\begin{tabular}{|l|l|l|l|}
\hline
\textbf{Network}               & \textbf{Alias} & \textbf{$|V|$} & \textbf{$|E|$} \\ \hline
moreno\_innovation              & $D_1$     & 241         & 1098\\ \hline
soc-ANU-residence             & $D_2$   & 217          & 2672 \\\hline
socfb-Caltech36            & $D_3$  & 769         & 16656\\ \hline
fb-pages-politician            & $D_4$  & 5908         & 41729\\ \hline
% soc-advogato                   & 6551         & 51332\\ \hline
\end{tabular}%
% %\vspace{-0.05in}
\end{table}

%\vspace{-0.2in}
\section{Experimental Results}
\label{sec:results}
% %\vspace{-0.05in}
% We implemented the proposed approach using SYCL.
% for heterogeneous computing. 
% Experiments on a many-core CPU utilized a dual 32-core AMD EPYC Rome 7452 with 64GB RAM, using Intel OneAPI SYCL and the \textit{icpx} compiler. GPU 
% The experiments were done on an Nvidia Ampere A100 with 80GB RAM, using Intel LLVM SYCL and the \textit{Clang++} compiler.
This section aims to showcase the effectiveness of our proposed algorithm for finding robust spanner (RS) nodes. The nodes identified as RS effectively span network communities, akin to those identified by existing spanner detection algorithms (see Section \ref{subsec:experiment_SpanningQuality}), while also retaining their crucial spanning capacity despite node failures (see Section \ref{sec:res_rob}). We analyze the scalability of the parallel implementation that lends itself to RS identification in large social networks undergoing constant evolution due to growth, failure, etc., (see Section \ref{subsec:results_GPU}). We implemented the proposed approach using CUDA. The experiments were conducted on an Nvidia Ampere A100 GPU equipped with 80GB of GPU memory and CUDA 11.8. 

\begin{table}[htbp!]
\caption{Average SHII comparison.}
%\vspace{-0.09in}
\centering
\begin{tabular}{|l|l|c|c|c|c|c|}
\hline
\textbf{Network} & \textbf{Diff.} & \textbf{RS} & \textbf{AP\_BICC} & \textbf{HAM} & \textbf{HIS} & \textbf{ABC}\\
\hline
$D_1$ & LT & 0.185 & 0.183 & 0.060 & 0.053 & \textbf{0.203} \\
 & IC & \textbf{0.502} & 0.490 & 0.071 & 0.076 & 0.415 \\ \hline
$D_2$ & LT & 0.240 & 0.271 & 0.215 & \textbf{0.348} & 0.407 \\
 & IC & 0.812 & 0.795 & 0.805 & \textbf{0.844} & 0.809 \\ \hline
% $D_3$ & LT & NA & NA & NA & NA \\
$D_3$ & IC & 0.807 & 0.850 & 0.800 & \textbf{0.864} & 0.853\\ \hline
$D_3'$ &LT &0.401	&0.407	&0.096	& \textbf{0.445} & 0.141\\ 
& IC &0.771	&0.831	&0.652	& \textbf{0.854} & 0.781\\ \hline
$D_4'$ &LT & \textbf{0.156}	&0.111	&0.060	&0.056 & 0.003\\
&IC & \textbf{0.795}	& 0.722	&0.528	&0.746 & 0.647\\
\hline
\end{tabular}
\label{table:SHII}
%\vspace{-0.1in}
\end{table}

\subsection{Spanning Quality Analysis}
\label{subsec:experiment_SpanningQuality}
We assess the spanning capability of a vertex by using the standard metric, \textit{structural hole influence index (SHII)}~\cite{he2016joint}.
% We utilize a metric called \textit{structural hole influence index (SHII)}~\cite{he2016joint} to measure the spanning quality of a vertex.
For a seed vertex $u_s$, 
%it is defined as 
$SHII(u_s) = \frac{(\sum_{c \in C_{all} \setminus C(u_s)}\sum_{v \in V_c}I(v))}{(\sum_{c \in C_{all}} \sum_{v \in V_c}I(v))}$, 
where $I(v)$ denotes the indication of a vertex $v$ being influenced under a specific information diffusion model, and $V_c$ is the set of vertices within community $c$. Thus, a higher value of SHII signifies better spanning capabilities.

To assess the spanning quality of the RS vertices, we utilize the networks listed in Table~\ref{table:graph_small}.  We calculate the average SHII for the top 25 vertices with high RSI and compare these with the top spanners identified through baselines like AP\_BICC~\cite{rezvani2015identifying}, HAM~\cite{he2016joint},  HIS~\cite{lou2013mining}, and Approximate Betweenness Centrality (ABC)~\cite{staudt2016networkit}. 
The SHII calculation employs two \textit{diffusion models}, Linear Threshold (LT) and Independent Cascade (IC)~\cite{kempe2003maximizing} for information dissemination. 

The SHII computation approach, as implemented in~\cite{gao2023easygraph}, struggles to complete within 24 hours on large graphs, yielding no results for dataset $D_4$ and partial results for $D_3$. 
To address this, we sample $D_3$ and $D_4$ and apply SHII on the sampled data. Our sampling strategy randomly selects community border vertices and incorporates their ego networks into the sampled datasets ($D_3'$ and $D_4'$), with a cap of approximately 2000 edges per sample.
The findings reported in Table~\ref{table:SHII} indicate that RS consistently achieves a medium to high SHII across datasets and diffusion models. 

%New Plots
\begin{figure*}[!ht]
% %\vspace{-0.1in}
	\centering
        % %\vspace{-0.1in}
        \subfloat{%
		\includegraphics[width=0.248\linewidth]{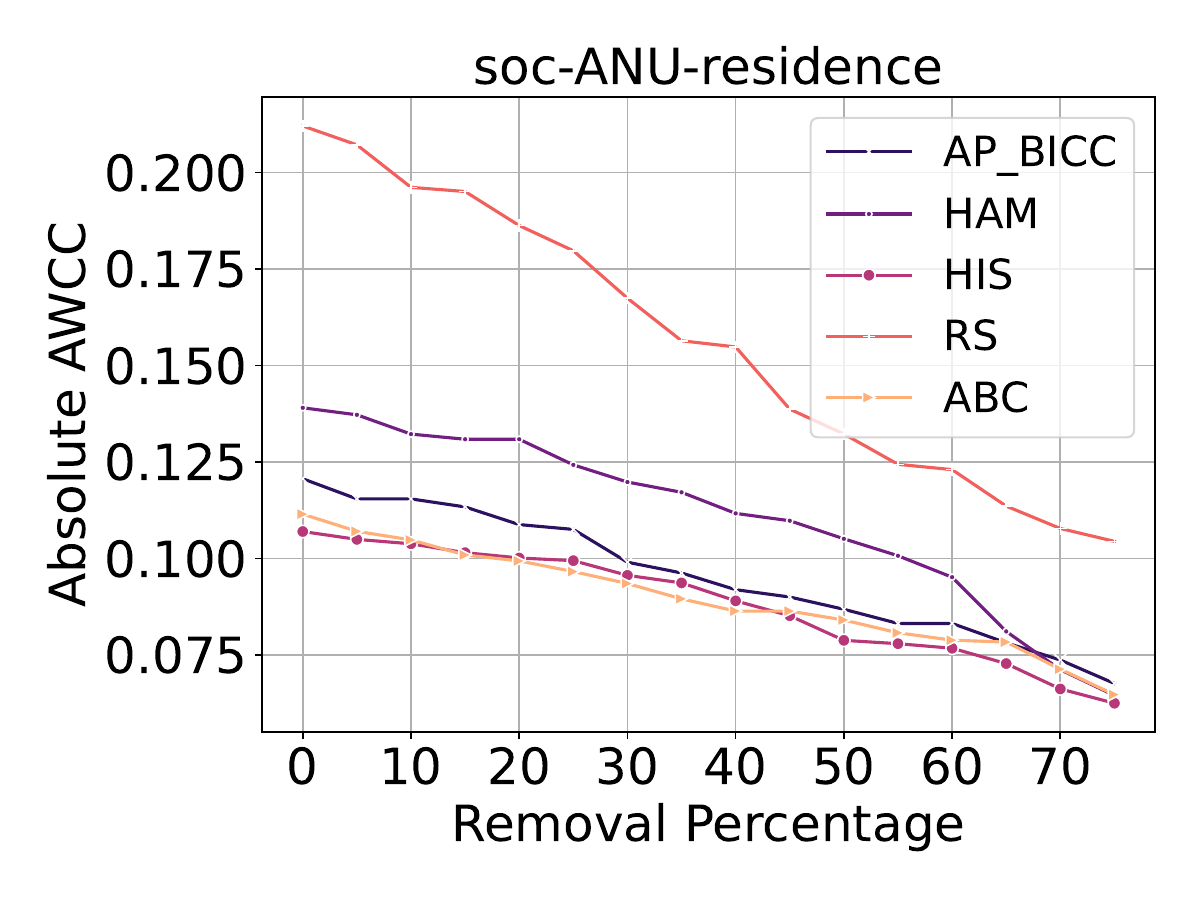}
		\label{}
	}\hspace{-0.1in}
        \subfloat{%
		\includegraphics[width=0.248\linewidth]{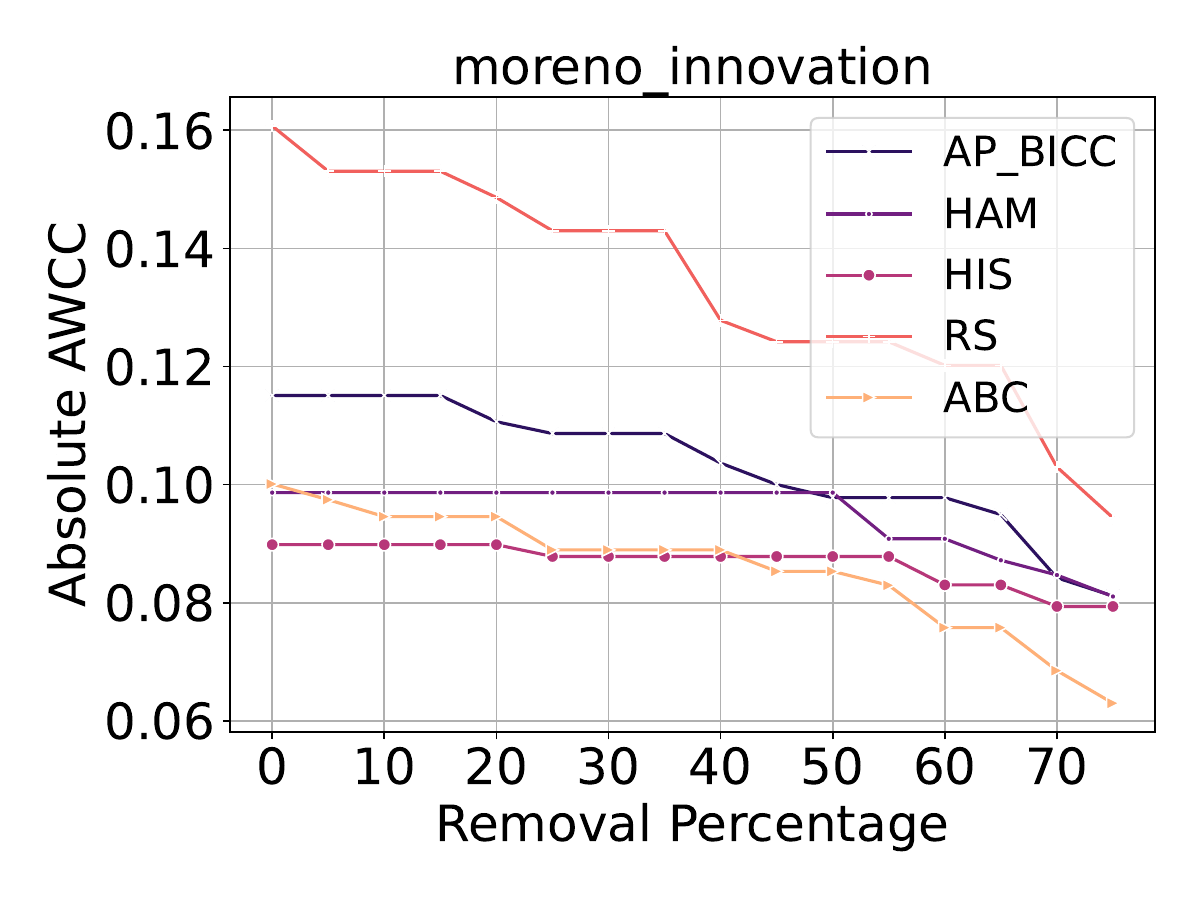}
		\label{}
	}\hspace{-0.1in}
        \subfloat{%
		\includegraphics[width=0.248\linewidth]{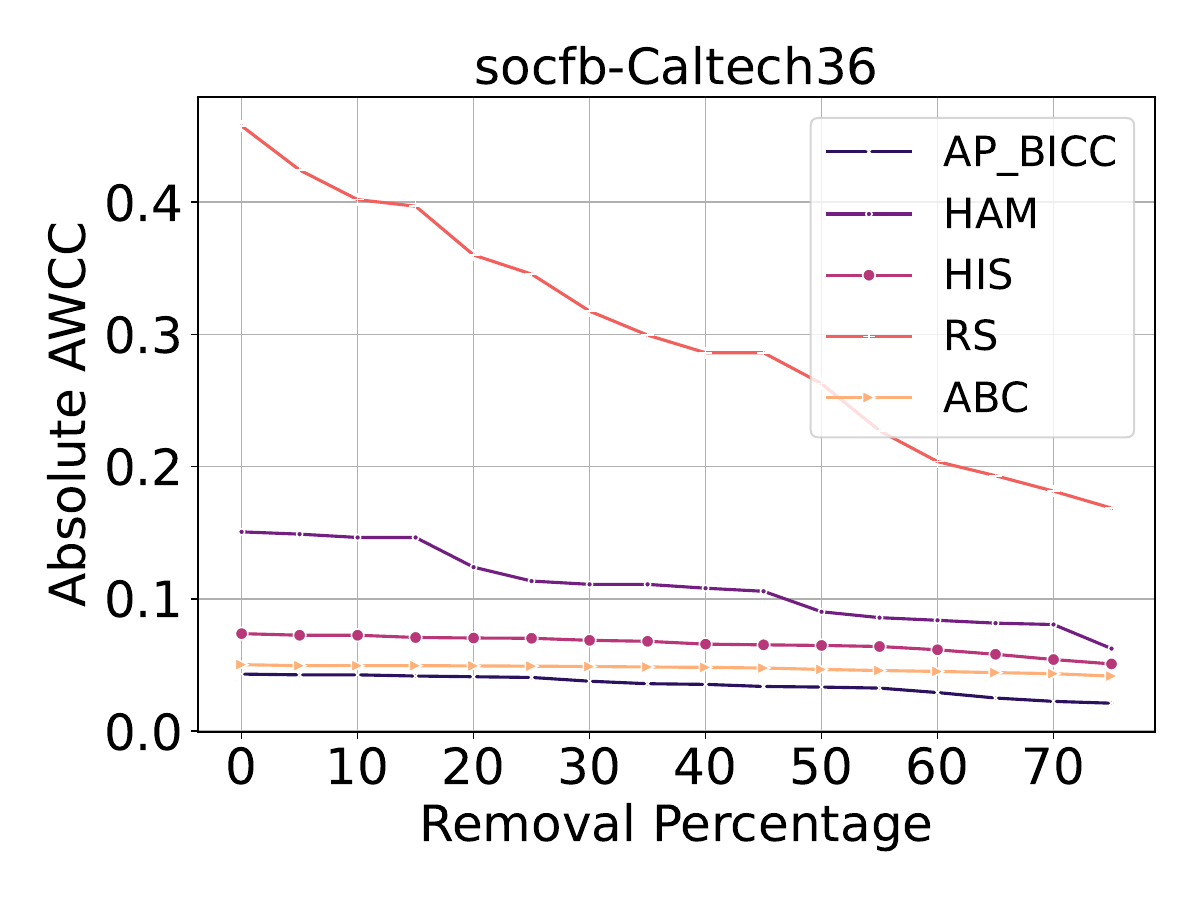}
		\label{}
	}\hspace{-0.1in}
        \subfloat{%
		\includegraphics[width=0.248\linewidth]{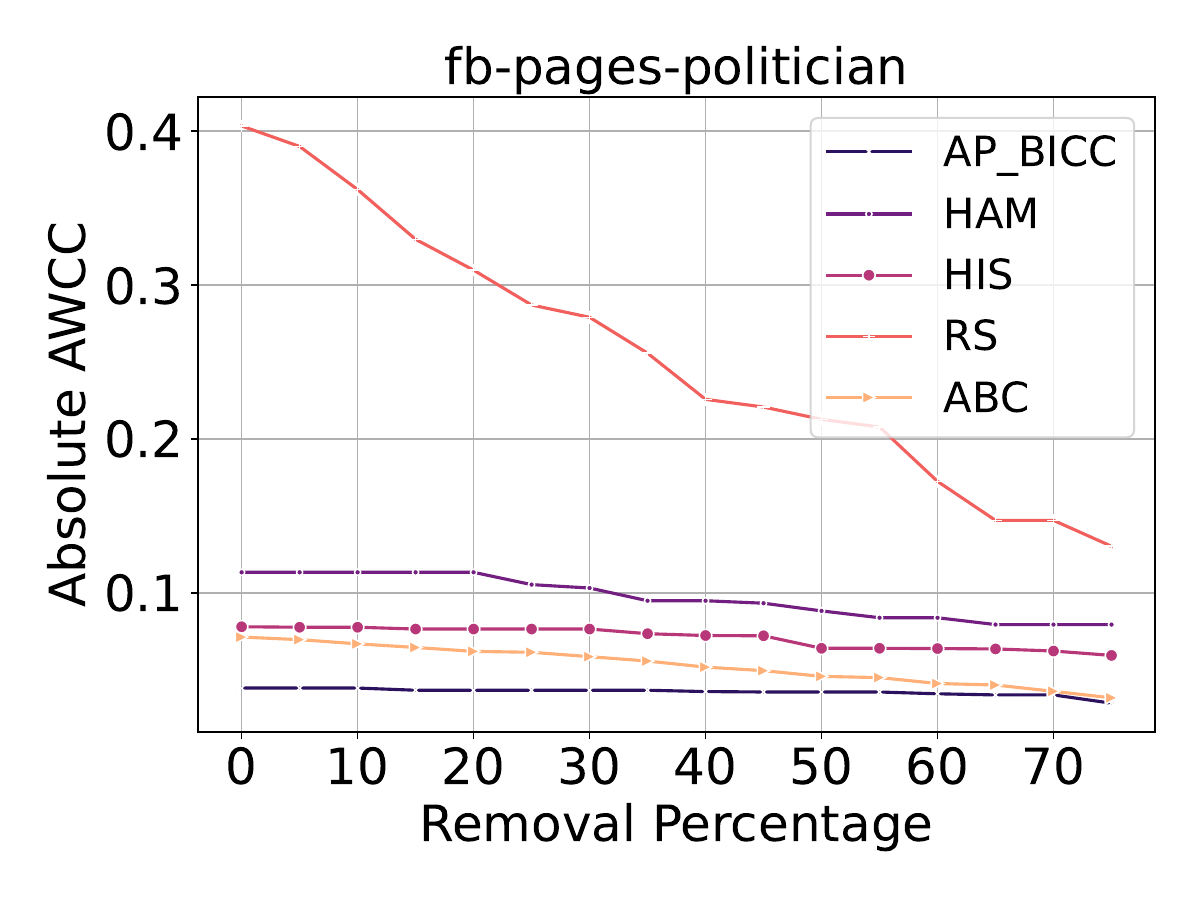}
		\label{}
	}
 \vspace{-0.15in}
	\caption{Impact of edge removal on community diversity of robust spanners.}
	\label{fig:robustness_analysis_edge_removal}
	%\vspace{-0.1in}
\end{figure*}

\begin{figure*}[htbp]
\vspace{-0.15in}
	\centering
        %\vspace{-0.05in}
        \subfloat{%
		\includegraphics[width=0.248\linewidth]{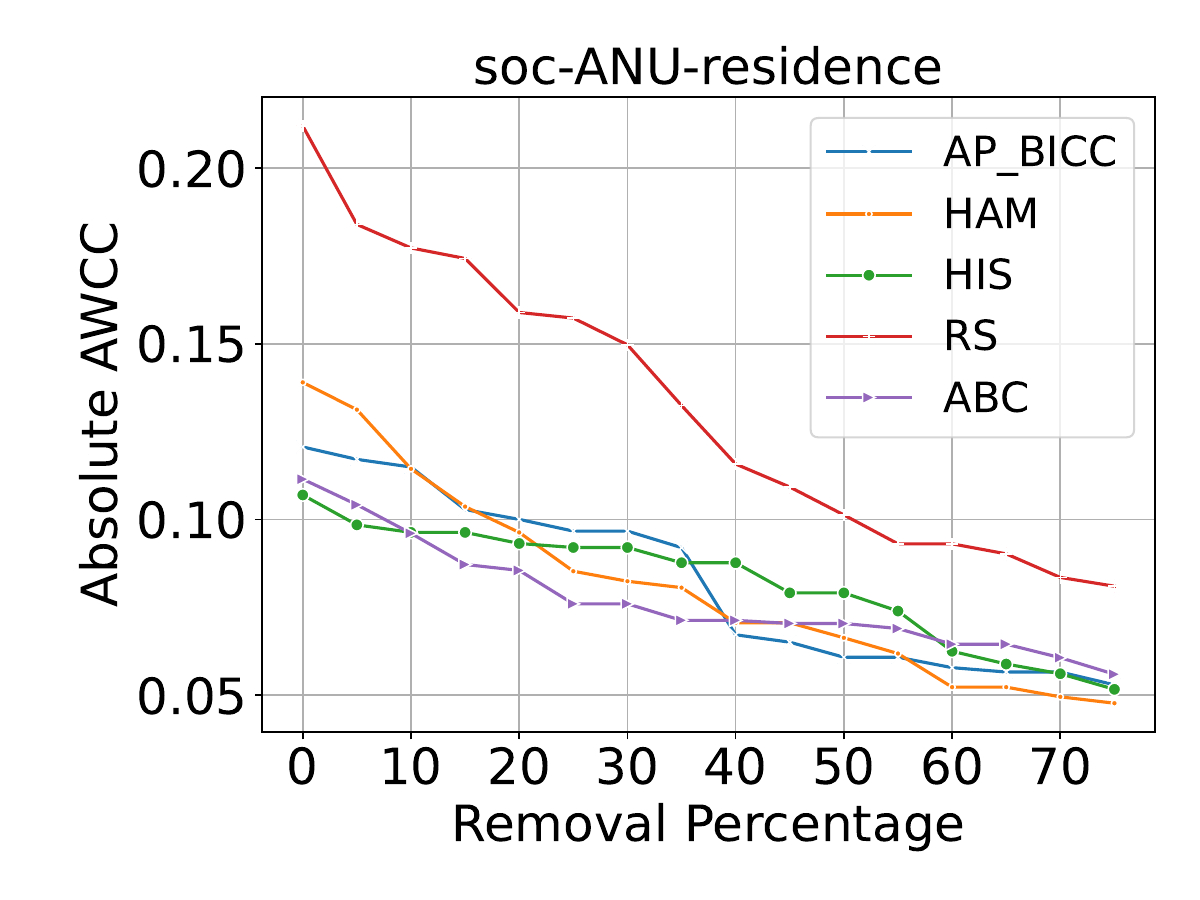}
		\label{}
	}\hspace{-0.1in}
        \subfloat{%
		\includegraphics[width=0.248\linewidth]{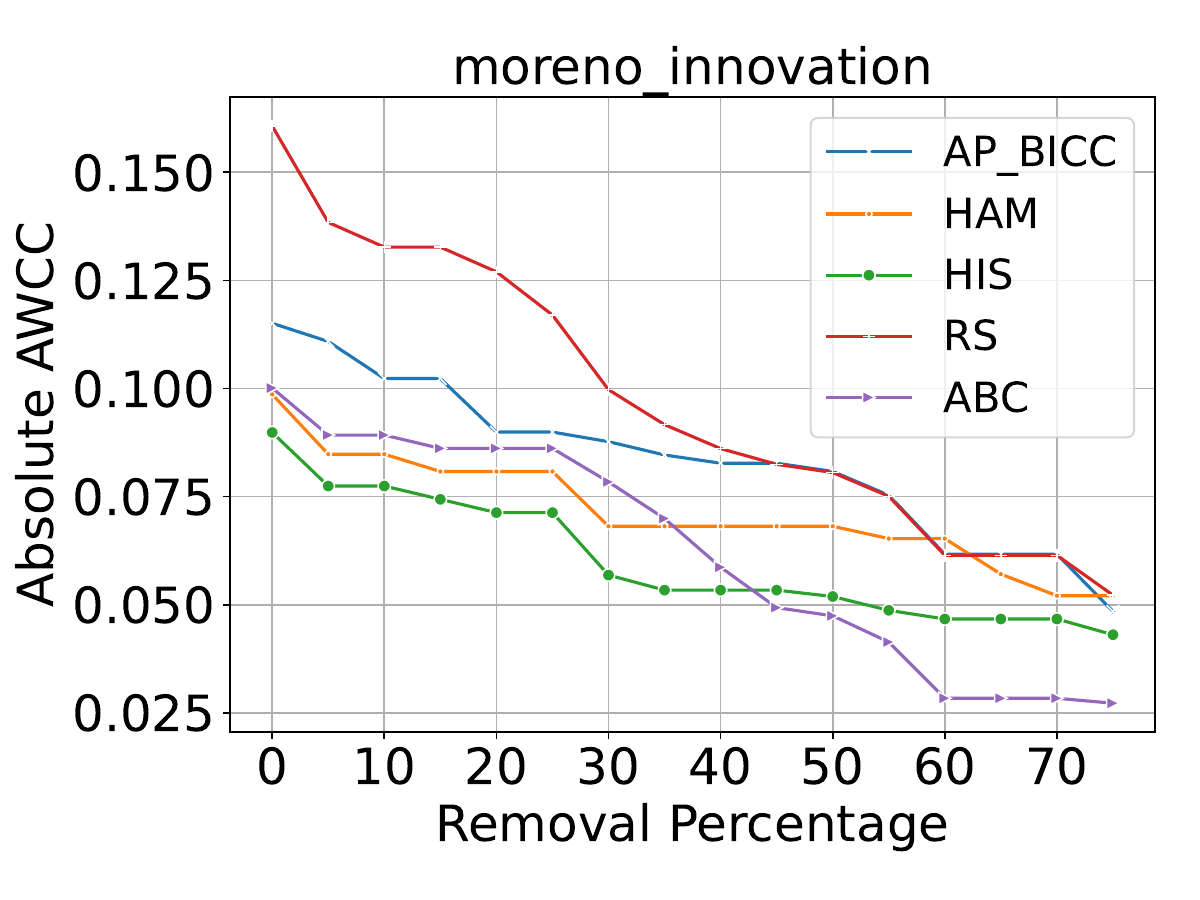}
		\label{}
	}\hspace{-0.1in}
        \subfloat{%
		\includegraphics[width=0.248\linewidth]{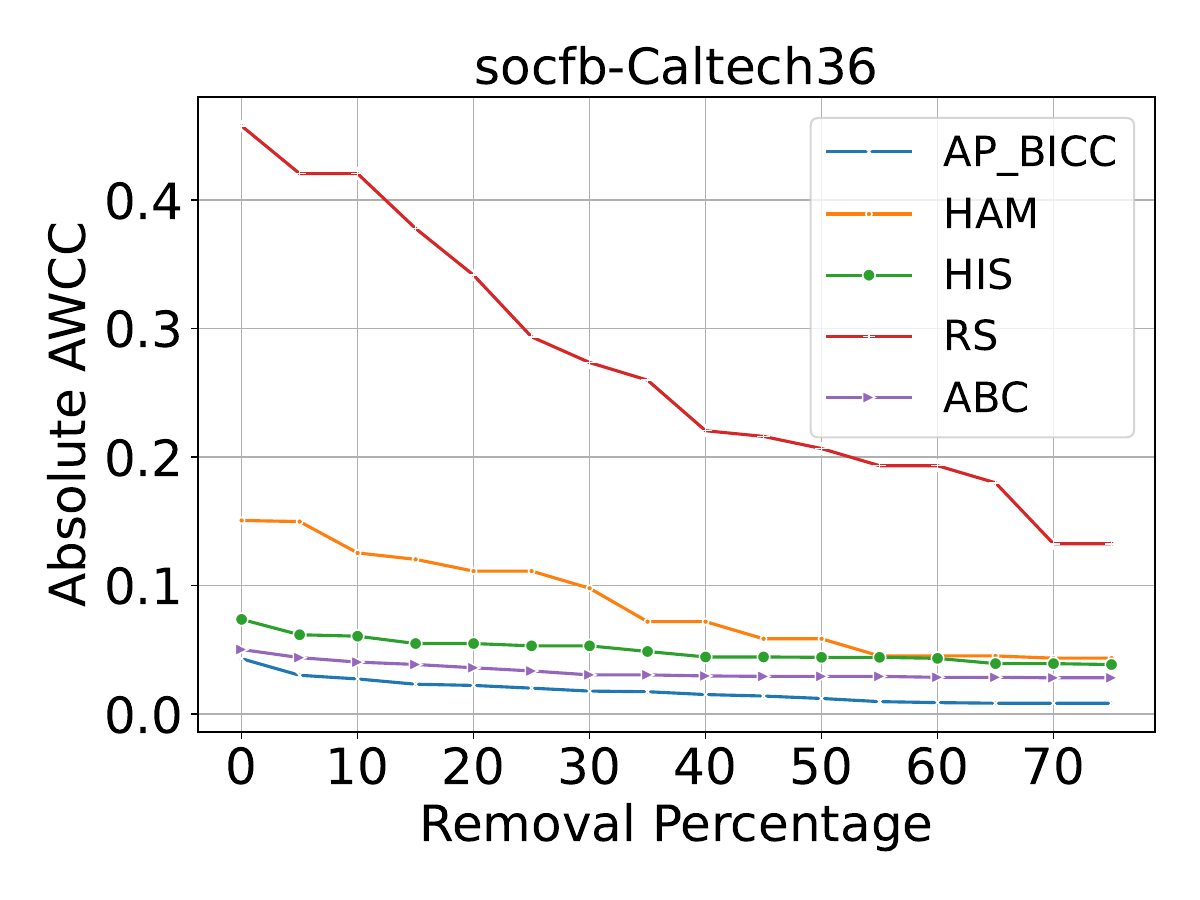}
		\label{}
	}\hspace{-0.1in}
        \subfloat{%
		\includegraphics[width=0.248\linewidth]{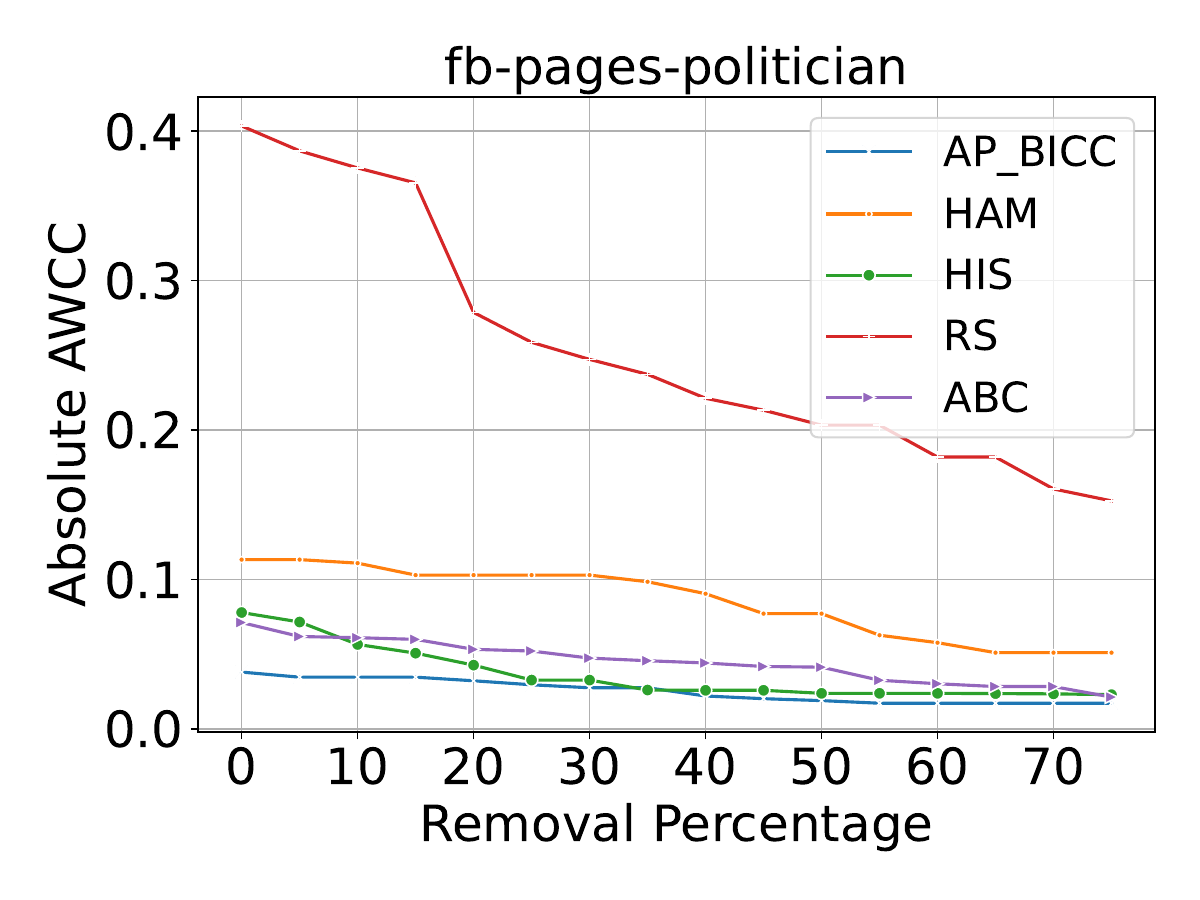}
		\label{}
	}
 \vspace{-0.1in}
	\caption{Impact of node removal on community diversity of robust spanners.}
	\label{fig:robustness_analysis_node_removal}
	%\vspace{-0.15in}
\end{figure*}

%\vspace{-0.1in}
\subsection{Robustness Analysis}\label{sec:res_rob}
%\textit{Network Perturbation Model: }
% %\vspace{-0.05in}
Information flow between communities can be hindered by factors such as communication congestion, and node or link failures. We simulate such disruptions by randomly removing nodes and links from social networks, followed by the evaluation of the robustness of the RS.
To quantify robustness, we use the \textit{average weighted number of connected communities} (AWCC) as a baseline metric for community-spanning diversity.
Given a set of vertices of interest $\mathcal{S}$, AWCC is defined as $\frac{1}{|\mathcal{S}|}\sum_{v \in \mathcal{S}}\frac{|\zeta (v)|}{d(v)}$, where $\zeta (v)$ denotes the set of community IDs of $v$’s neighbors~\cite{rezvani2015identifying, li2019distributed, khanda2024structural}. Specifically, we calculate the impact of disruption on the AWCC of $\mathcal{S}$, by recalculating the reachable communities while accounting for paths rendered unusable by the network disruption. Specifically, we use a modified version of the AWCC metric, referred to as \textit{absolute} AWCC, in which only $|\zeta (v)|$ is altered based on the network disruption.

This experiment utilizes four social network datasets (Table~\ref{table:graph_small}) and compares the robustness of the top 25 spanners detected by our method with those by various baseline algorithms, including AP\_BICC~\cite{rezvani2015identifying}, HAM~\cite{he2016joint}, HIS~\cite{lou2013mining}, and approximate betweenness centrality (ABC)~\cite{staudt2016networkit}. 
Fig.~\ref{fig:robustness_analysis_edge_removal} and ~\ref{fig:robustness_analysis_node_removal} show the performance of spanners obtained by different methods in terms of absolute AWCC following the removal of edges and vertices, respectively. The experiment begins with the original networks and progresses with the random removal of 5\% of the edges/nodes at each iteration. This process continues until up to 75\% of edges/nodes are removed. The absolute AWCC of the spanner sets is computed at each iteration. As the deletion of edges/nodes may disrupt the connectivity among communities, the AWCC score tends to decrease gradually with an increase in the percentage of edge and node removal. For all cases, the RS vertices maintain the highest AWCC score, indicating superior robustness in the presence of network disruptions.

\begin{table}[htb]
% %\vspace{-0.12in}
\caption{Social Networks Dataset (Large)}
%\vspace{-0.05in}
\label{table:graph_larg}
\centering
% \resizebox{\textwidth}{!}{%
\begin{tabular}{|l|l|l|}
\hline
\textbf{Network}               & \textbf{$|V|$} & \textbf{$|E|$} \\ \hline
Coauthor~\cite{lou2013mining}             & 53,442         & 127,968 \\\hline
Twitter~\cite{lou2013mining}             & 92,180         & 188,971\\ \hline
soc-lastfm~\cite{nr}           & 1,191,805         & 4,519,330\\ \hline
soc-pokec~\cite{nr}   & 1,632,803   & 22,301,964\\ \hline
Patent~\cite{lou2013mining}            & 2,445,351         & 5,804,865\\ \hline
soc-orkut~\cite{nr}   & 2,997,166   & 106,349,209\\ \hline
soc-livejournal~\cite{nr} & 4,033,137   & 27,933,062\\ \hline
soc-sinaweibo~\cite{nr}   & 58,655,849    & 261,321,071\\ \hline
\end{tabular}%
% %\vspace{-0.1in}
\end{table}

% %\vspace{-0.07in}
\subsection{Execution Time Analysis}
\label{subsec:results_GPU}
Here we conduct experiments on eight large datasets (See Table~\ref{table:graph_larg}). Community ground truth for \textit{coauthor}, \textit{twitter}, and \textit{patent} graphs is taken from~\cite{lou2013mining}, while for the remaining graphs, communities are generated using the Louvain method~\cite{staudt2016networkit}. As the concept of RS is a new contribution, there is no existing direct baseline for comparison. 
Thus, we evaluate the execution time of our GPU-based implementation against the most relevant SHS detection approaches.
% and a parallel betweenness centrality computation.
To the best of our knowledge, ESH~\cite{li2019distributed} is the only parallel SHS detection algorithm designed on the PowerGraph framework in a distributed computing environment. However, it lacks an open-source parallel implementation for direct comparison with our GPU-based approach.
Contrasting with our RS detection, the other benchmark, HAM~\cite{he2016joint} does not rely on known community IDs, but jointly identifies communities and SHS. Similarly, AP\_BICC~\cite{rezvani2015identifying} does not leverage community ground truth, opting to calculate vertices' bounded inverse closeness centralities to detect SHS instead. HIS~\cite{lou2013mining}, akin to our method, utilizes community structures to pinpoint SHS within user-specified target communities. 
As reported in Table~\ref{table:SHII}, HIS produces spanners with high spanning quality. 
In the absence of prior work on robust spanners or parallel SHS detection, we benchmark our method against the sequential HIS.
% , which is most closely aligned with our approach.

\begin{figure*}[htbp]
% %\vspace{-0.1in}
	\centering
       \vspace{-0.15in}
        \subfloat{%	
        \includegraphics[width=0.14\linewidth]{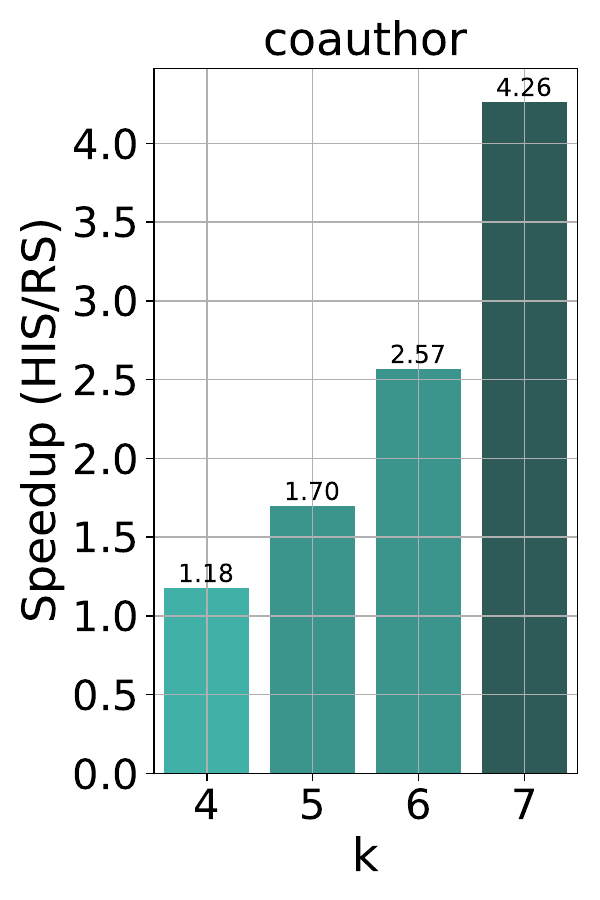}
		% \label{fig:eg2}
	}
        \hspace{-0.1in}
        \subfloat{%
	%%\vspace{-0.1in}	
 \includegraphics[width=0.14\linewidth]{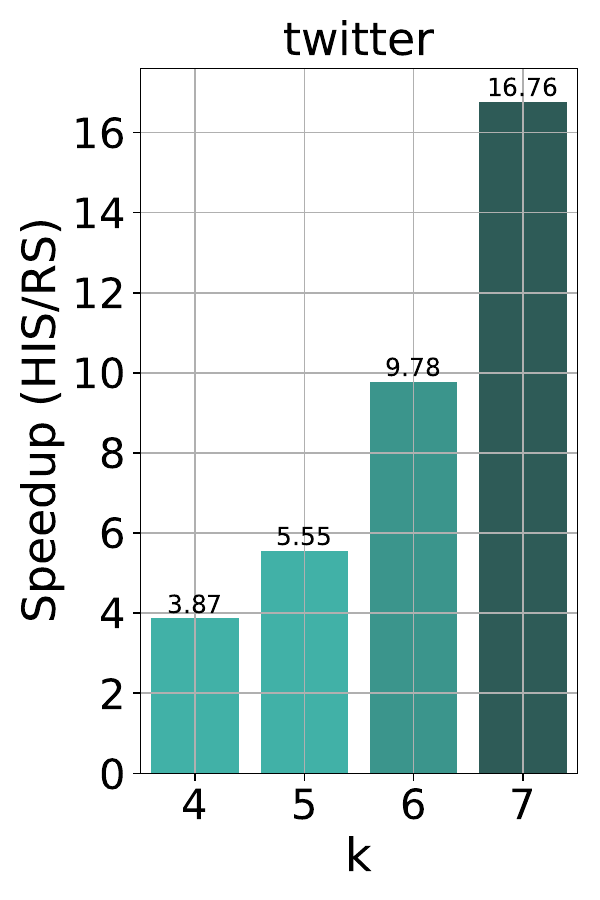}
		% \label{fig:eg1}
	}
        \hspace{-0.1in}
        %%\vspace{-0.1in}
        \subfloat{%
		\includegraphics[width=0.14\linewidth]{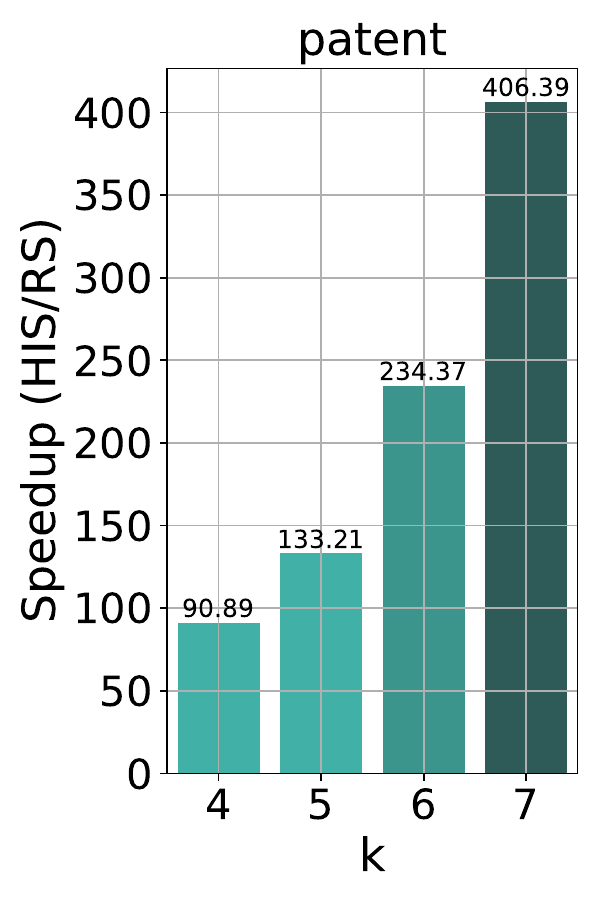}
		% \label{fig:eg2}
	}
        \hspace{-0.1in}
        \subfloat{%
		\includegraphics[width=0.14\linewidth]{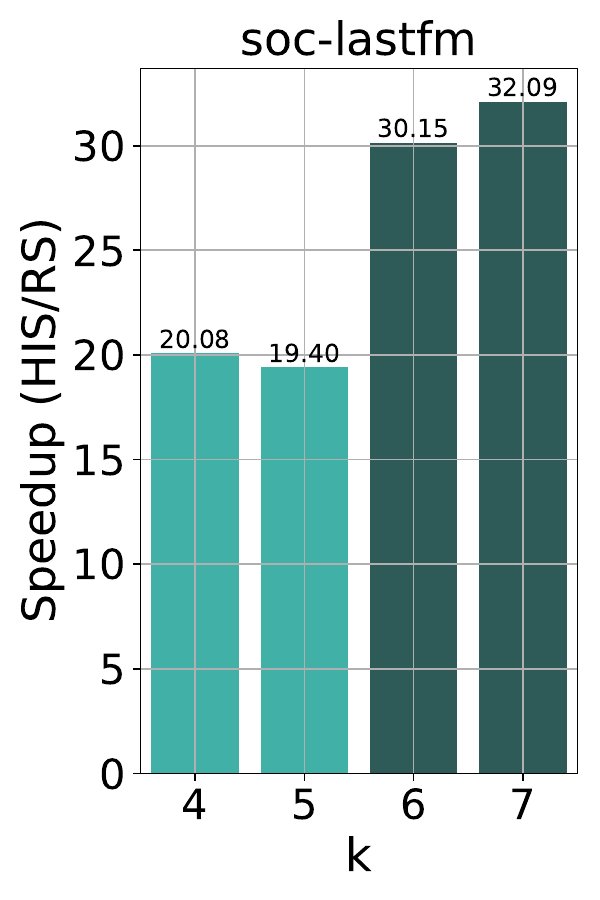}
		% \label{fig:eg2}
	}
        \hspace{-0.1in}
        \subfloat{%
		\includegraphics[width=0.14\linewidth]{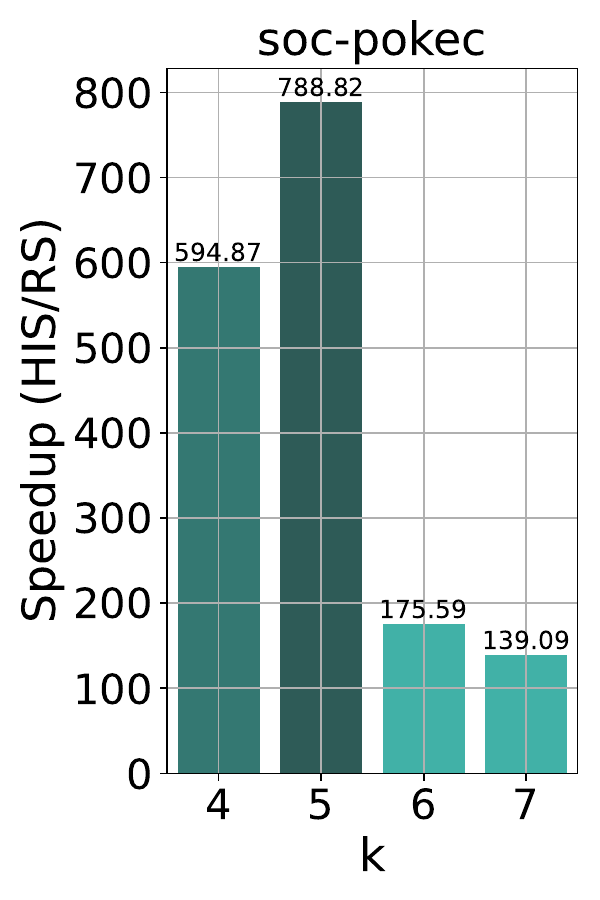}
		% \label{fig:eg2}
	}
        \hspace{-0.1in}
        \subfloat{%
		\includegraphics[width=0.14\linewidth]{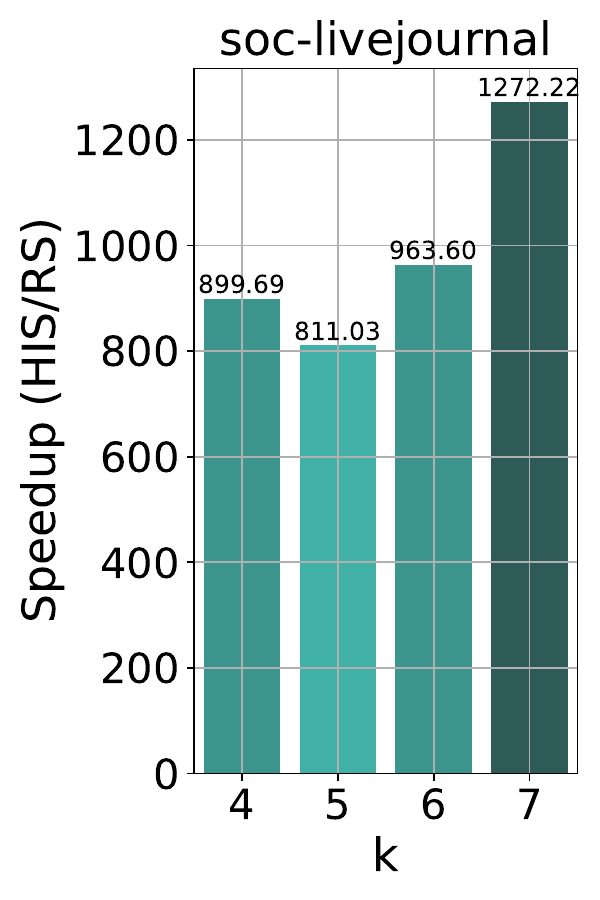}
		% \label{fig:eg2}
	}
        \hspace{-0.1in}
        \subfloat{%
		\includegraphics[width=0.14\linewidth]{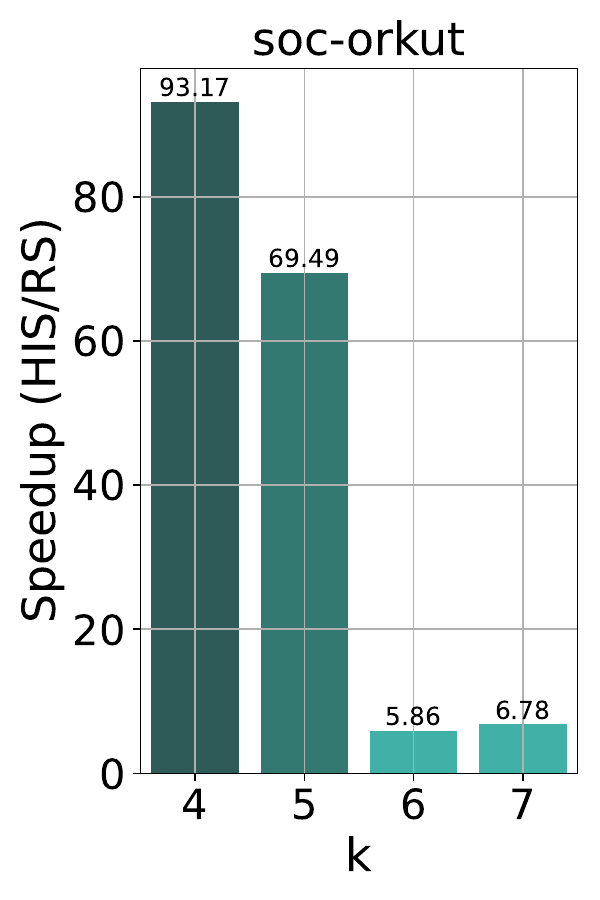}
		% \label{fig:eg2}
	}
        
 \vspace{-0.15in}
	\caption{Speedup comparison with HIS. $k$ denotes the total targeted communities}
	\label{fig:GPU_and_HIS_analysis}
	\vspace{-0.1in}
\end{figure*}

\begin{figure*}[htbp]
 %\vspace{-0.15in}
	\centering
       \vspace{-0.15in}
        \subfloat{%	
        \includegraphics[width=0.14\linewidth]{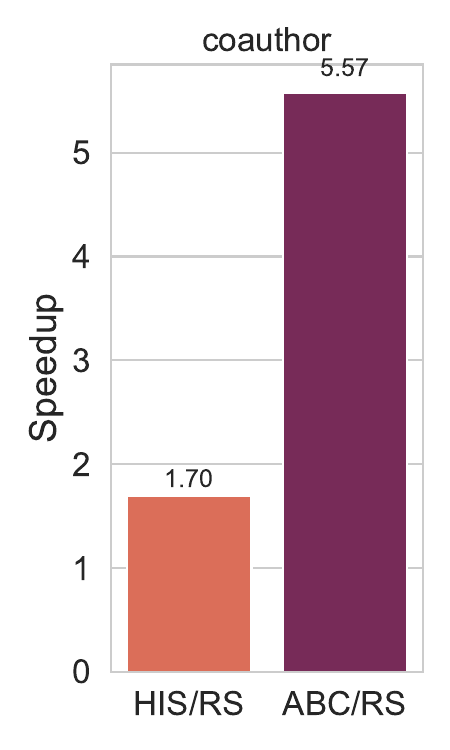}
		% \label{fig:eg2}
	}
        \hspace{-0.1in}
        \subfloat{%
	%%\vspace{-0.1in}	
 \includegraphics[width=0.14\linewidth]{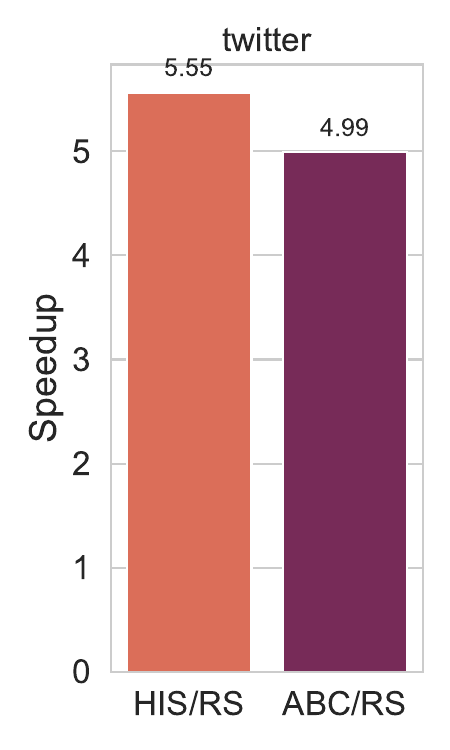}
		% \label{fig:eg1}
	}
        \hspace{-0.1in}
        %%\vspace{-0.1in}
        \subfloat{%
		\includegraphics[width=0.14\linewidth]{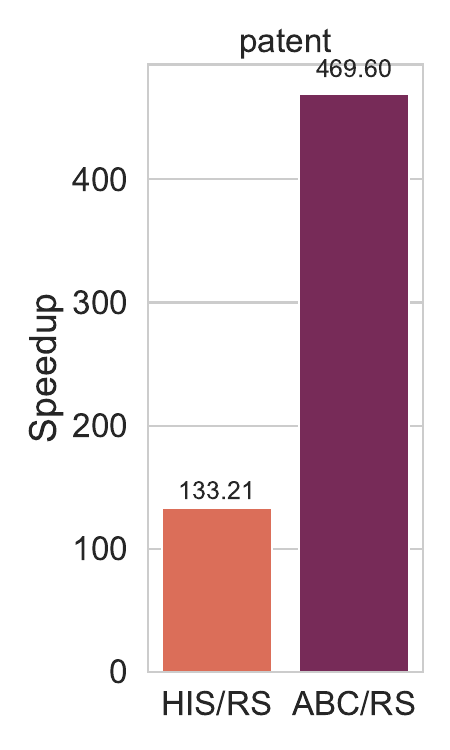}
		% \label{fig:eg2}
	}
        \hspace{-0.1in}
        \subfloat{%
		\includegraphics[width=0.14\linewidth]{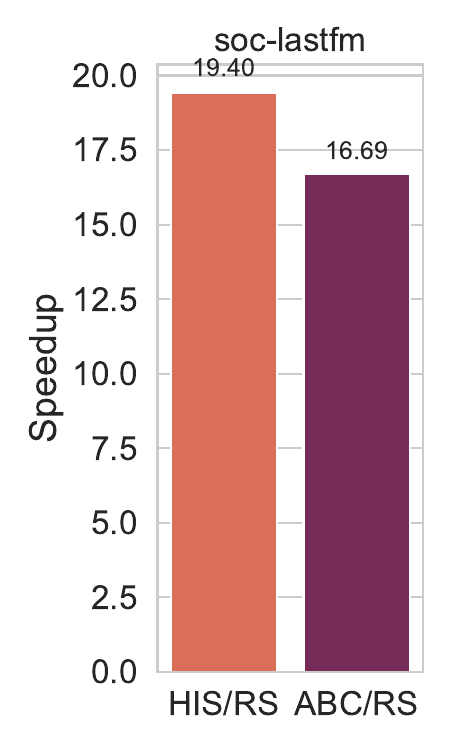}
		% \label{fig:eg2}
	}
        \hspace{-0.1in}
        \subfloat{%
		\includegraphics[width=0.14\linewidth]{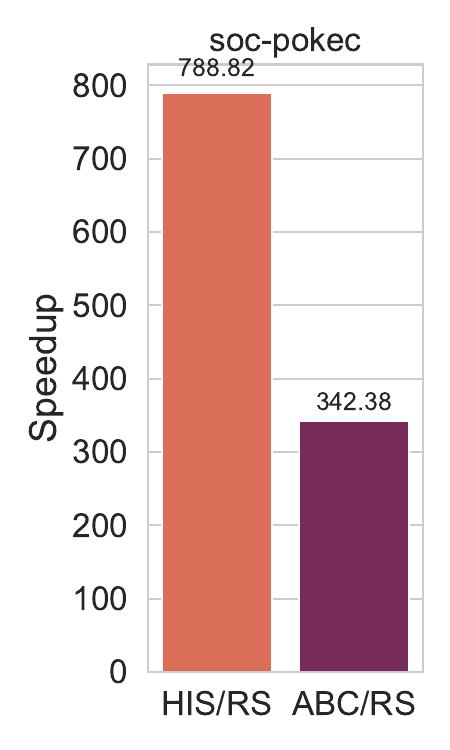}
		% \label{fig:eg2}
	}
        \hspace{-0.1in}
        \subfloat{%
		\includegraphics[width=0.14\linewidth]{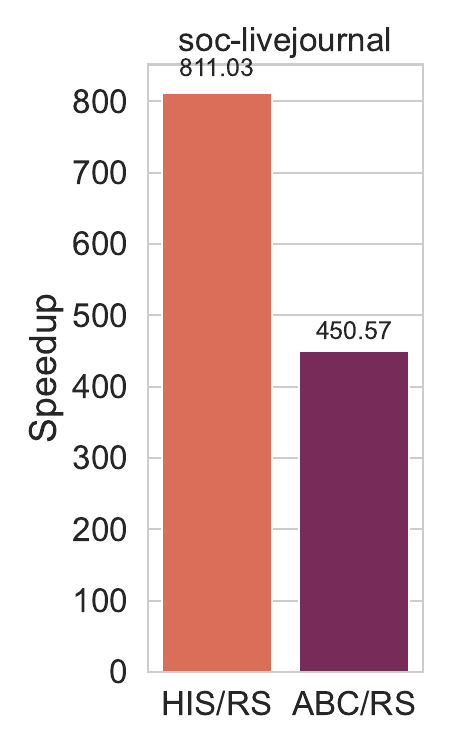}
		% \label{fig:eg2}
	}
        \hspace{-0.1in}
        \subfloat{%
		\includegraphics[width=0.14\linewidth]{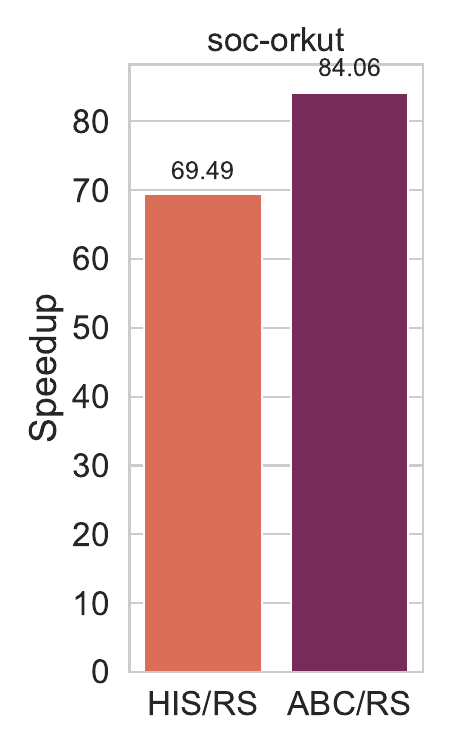}
		% \label{fig:eg2}
	}
        
 \vspace{-0.15in}
	\caption{Speedup of our implementation over HIS and ABC.}
	\label{fig:speedup2}
	\vspace{-0.1in}
\end{figure*}

 \begin{figure}[htp]
    \vspace{-0.2in}
    \centering \includegraphics[width=0.9\linewidth]{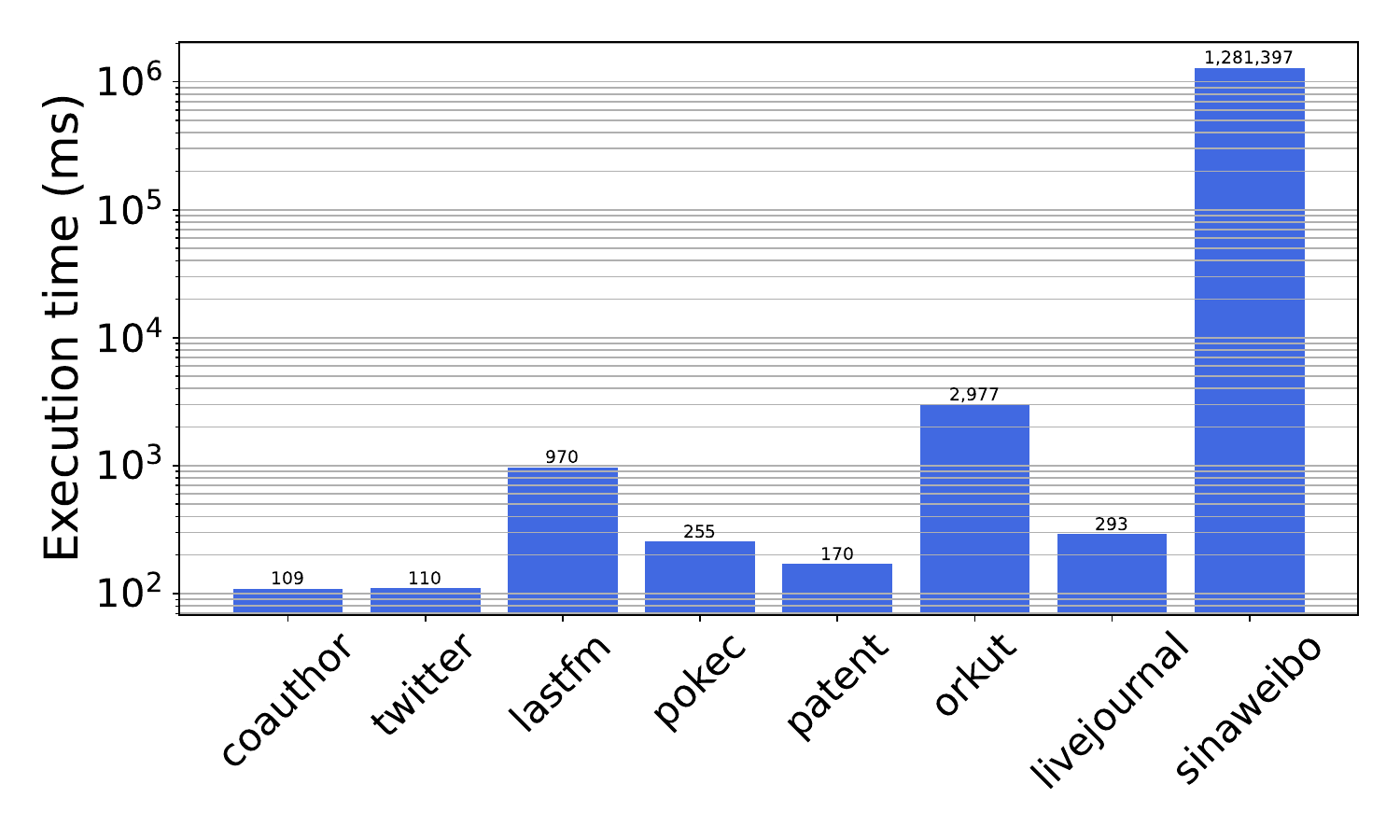}
    \vspace{-0.15in}
    \caption{ Total execution time analysis.
    }
    \label{fig:timeplot}
    \vspace{-0.1in}
\end{figure}

\begin{figure}[htp]
    \vspace{-0.1in}
    \centering \includegraphics[width=0.9\linewidth]{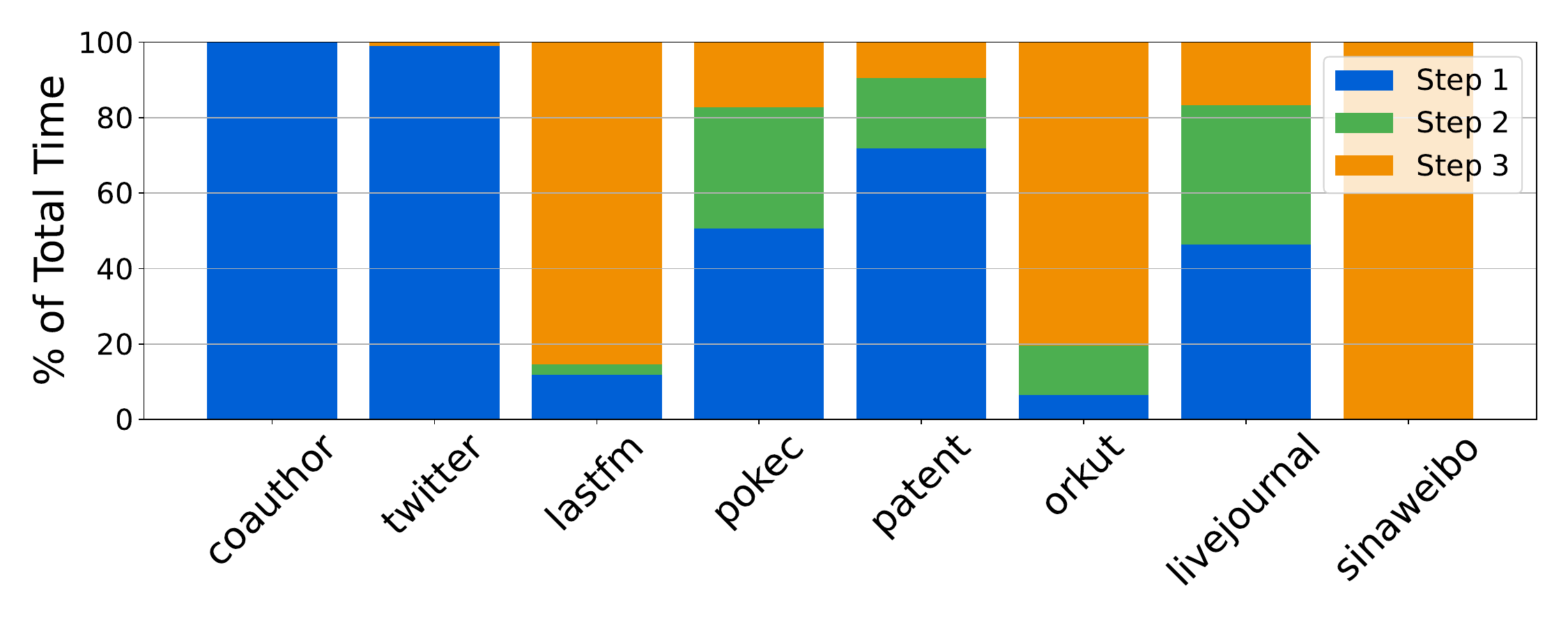}
    \vspace{-0.15in}
    \caption{ Stepwise execution time percentage analysis.
    }
    \label{fig:timesteps}
    \vspace{-0.1in}
\end{figure}

In our first execution time experiment, we evaluate \textit{speedup}, defined as the ratio of the execution time of the most effective and closest sequential baseline, HIS, to that of our RS implementation.
In both HIS and RS implementations, the target communities for spanner detection are provided as input. In our experiments, we sort communities in descending order by their number of member vertices and select the top $k$ for evaluation. Since the top 7 communities typically cover a large portion of each network, we vary $k$ from 4 to 7.
Fig.~\ref{fig:GPU_and_HIS_analysis} shows that RS consistently outperforms HIS across all datasets and $k$ values, achieving speedups ranging from $1.18 \times$ to $1272 \times$, with an average speedup of $244 \times$.
Speedup generally increases with larger $k$ values, as HIS’s execution time grows more rapidly than that of RS. However, an exception is observed in \textit{soc-pokec} and \textit{soc-orkut}, where speedup declines beyond $k=5$. This is due to a sharp rise in RS execution time, attributed to a significant increase in valid triads. While we systematically select the top $k$ communities based on size, the actual triad formation depends on the underlying community structure, making predicting workload increases from adding additional communities challenging.

As discussed in Section~\ref{subsec:experiment_SpanningQuality}, in addition to SHS detection methods, approximate betweenness centrality (ABC) is used to assess spanning quality. Due to the lack of parallel SHS implementations, we adopt a shared-memory parallel version of ABC~\cite{staudt2016networkit} as a secondary baseline. With $k=5$ for both HIS and RS, and using 64 CPU threads for ABC, Fig.~\ref{fig:speedup2} presents RS speedups over HIS and ABC across seven large social networks (excluding \textit{soc-sinaweibo}, where HIS and ABC failed to execute). RS achieves speedups over ABC ranging from $4.99\times$ to $811\times$, with an average of $196\times$.

Fig.~\ref{fig:timeplot} presents the total execution time of RS across all eight large graphs, including the time required to transfer the input graph from the CPU to the GPU and to transfer the results back to the CPU. As the largest graph, \textit{sinaweibo} incurs the highest execution time of approximately $21$ minutes, while the smaller graphs \textit{coauthor} and \textit{twitter} complete in about $110$ milliseconds. Execution time increases with graph density due to a greater likelihood of valid triad formation. For example, although \textit{patent} has a similar number of vertices as \textit{orkut}, and \textit{livejournal} has 1.4 times more vertices, \textit{orkut} is significantly denser and requires more processing time.

To further analyze the execution time, Fig.~\ref{fig:timesteps} shows the percentage of total time spent in the three steps of the RS algorithm. As observed earlier, speedup is lower for smaller graphs such as \textit{coauthor} and \textit{twitter}, primarily because Step 1, which identifies border vertices after transferring the graph from CPU to GPU, dominates the execution time. This indicates that fixed overheads are substantial compared to the actual computation, limiting the achievable speedup in these cases.
For high-runtime graphs such as \textit{lastfm}, \textit{orkut}, and \textit{sinaweibo}, Step 3 accounts for the largest share of execution time, reflecting the high computational cost of triad detection.
A straightforward scalability experiment is not feasible here, as the number of valid triads depends on both graph size and the underlying community structure. As a result, understanding total workload with varying vertices, edges, or communities remains a complex problem and requires further research.

\section{Related Works}
\label{sec:refs}
% This section reviews the existing methods to identify SHS and methods to measure robustness.
%\vspace{-0.05in}
\subsection{Structural Hole Spanner Identification}
Structural hole spanner (SHS) detection algorithms can be broadly divided into three major categories. However, a significant drawback of SHS detection is its NP-hardness nature, resulting in high computational costs and ineffective scalability for large networks. 

%Three major categories can be used to group structural hole spanner detection techniques.  However, the NP-hardness of the problem \cite{rezvani2015identifying, song2015mining} is the main drawback of current SHS detection techniques, resulting in high computational costs and ineffective scalability for large networks.

%****important summary
% Methods falling within this category concentrate on recognizing spanners that connect various communities in the network~\cite{lou2013mining,tang2012inferring,wang2011detecting,katz1964personal, wang2011detecting}. The subsequent category employs measures of network centrality to locate SHS. This involves considering a node's position and significance within a social network~\cite{granovetter1973strength, li2019distributed, song2015mining, rezvani2015identifying}.

The first category focuses on information distribution and aims to identify SHS nodes by evaluating the impact of their removal on network connectivity~\cite{lou2013mining, tang2012inferring}. Intuitively, the  nodes whose removal causes greater disruption to the information flow are more likely to be SHS.
% Similarly, in ~\cite{wang2011detecting}, a semi-supervised link prediction technique is developed to predict the types of social relationships across heterogeneous networks. 
% Note that both the above methods require community labels of the nodes in advance. 
% In ~\cite{wang2011detecting} a robust hierarchical overlapping community detection framework using Personalized PageRank (PPR) is proposed. Based on the community structures, they further proposed a breadth-first search approach to identify top-$K$ SHS.   
Two methodologies HIS and MaxD introduced in~\cite{lou2013mining} target the identification of the top $K$ SHS nodes in large social networks. 
% Based on the premise that SHS exhibit a heightened propensity to establish connections with opinion leaders~\cite{katz1964personal}, they formulated the concepts of importance score and SHS score for individual nodes. 
Using the community labels of the nodes, the HIS algorithm assigns higher scores to nodes that receive significant information from neighboring nodes, reflecting their potential as SHS. In contrast, MaxD seeks to identify the top $K$ nodes whose removal maximally reduces the minimum cut of a community, effectively fragmenting the network. While both algorithms can detect quality spanners, HIS is computationally more efficient than MaxD. In~\cite{tang2012inferring}, the authors introduced a method called the Transfer-based Factor Graph (TranFG) model, which assumes that SHS nodes tend to form links with kernel members from different communities. Both of these methods \cite{lou2013mining,tang2012inferring} require the community labels as input, which is challenging to obtain in a large-scale network.
% This approach effectively identifies the leading $K$ SHS. Notably, HIS and MaxD methodologies exhibit superior performance in comparison to both community kernel detection~\cite{wang2011detecting} and link prediction~\cite{tang2012inferring} techniques for SHS identification. 
%The kernel detection approach is founded upon the assumption that SHS tend to establish links with kernel members from divergent communities. Conversely, the link prediction technique is developed to predict the types of social relationships with the help of SH analysis. However, these methods \cite{wang2011detecting,tang2012inferring} require community labels of nodes in advance.
The authors in ~\cite{li2019distributed} have introduced the ESH algorithm that uses a factor diffusion process in the PowerGraph distributed parallel graph processing framework to detect SHS in a large social network. The factor diffusion process captures the diversity of the information flow to identify SHS nodes.

The second category of algorithms focuses on network centrality, i.e., evaluating whether nodes occupy strategically advantageous positions within social networks. For example, the authors in~\cite{song2015mining} proposed a heuristic to detect the top $K$ SHS suggesting that a higher number of weak ties significantly increases the likelihood that a node is an SHS.
In ~\cite{liao2020detecting}, the authors proposed a conductance–degree SH detection algorithm, enhanced by an improved label propagation technique. This approach combines conductance and degree scores to identify SHS nodes while effectively filtering out irrelevant nodes. Similarly, AP\_BICC~\cite{rezvani2015identifying} leverages the bounded inverse closeness centrality and articulation points to detect SHS. This method is based on the premise that removing SHS nodes would lead to a substantial increase in the average shortest path length within the resulting subgraph.

Another line of research introduces the Harmonic Modularity (HAM) method ~\cite{he2016joint}, which simultaneously detects communities and structural hole spanners in networks using only the topological information. HAM employs harmonic function analysis to compute harmonic modularity for community identification, while also highlighting nodes that connect multiple communities as potential SHS candidates.
Further, in ~\cite{he2021understanding}, the authors conducted a comprehensive analysis of SHS behavioral traits across demographic, spatiotemporal, and linguistic dimensions within Location-Based Social Networks (LBSNs), using data from Foursquare. 
% Leveraging these insights, 
They developed a supervised learning model to distinguish between SHS and regular users. Similarly, ~\cite{gong2019identifying} proposes a classifier to pinpoint SHS in online social networks by leveraging users' profiles and user-generated contents (UGCs), thus eliminating the need for the entire social graph.
% In ~\cite{he2021understanding}, the authors first conduct a comprehensive analysis of the behavior characteristics of SHS in demographic, spatiotemporal, and linguistic aspects in Location-Based Social Networks (LBSNs) using data from Foursquare. Based on these characteristics, they propose a supervised machine learning-based prediction model to accurately distinguish between SHS and ordinary users. In ~\cite{gong2019identifying}, a similar supervised machine learning-based classifier model is proposed to identify SHS in online social networks.
% without the need for the entire social graph, leveraging users' profiles and user-generated contents (UGCs). The model further utilizes ego networks and a cross-site linking function to enhance identification performance. 
% ~\cite{nie2022joint} introduces a novel algorithm, SDHE, for simultaneously detecting communities and structural hole spanners (SHS) in networks using hyperbolic embedding. 
% The approach embeds scale-free networks into hyperbolic space, utilizing the critical gap and angular distribution of nodes to identify communities and SHS. However, in the proposed approach, computational complexity is highly impacted by the hyperbolic embedding algorithm used.
In~\cite{nie2022joint}, the authors first embedded scale-free networks into hyperbolic space and then utilized the critical gap and angular distribution of nodes for identification of communities and SHS simultaneously. However, the efficiency of this method largely depends on the choice of a hyperbolic embedding algorithm. 

%\vspace{-0.1in}
\subsection{Robustness Metrics}
% %\vspace{-0.05in}
Network robustness is generally refers to a network's ability to maintain its structure and functionality despite failures~\cite{dekker2004network}. In this context, structure and function are deeply interconnected: alterations in structural attributes, such as connectivity and path length can significantly impact functional attributes like data throughput and communication latency, and vice versa~\cite{blanchini2011structurally}. Multiple metrics have been proposed to evaluate robustness from different perspectives. \textit{Distance-based} metrics, including characteristic path length, global and local network efficiency, shortest path length, and diameter, measure the length and number of paths connecting nodes pairs. They emphasize the importance of maintaining shorter paths to enable faster  data transmission in fewer hops~\cite{xing2013reformulation,oehlers2021graph}.

Complementing distance-based metrics are \textit{percolation-based} connectivity metrics~\cite{callaway2000network}, which examine a network's resilience to fragmentation under a multitude of failure scenarios. These include vertex and edge connectivity, measuring the minimum number of nodes or edges whose removal would disconnect the network; the \textit{reliability polynomial}, which estimates the probability of maintaining network connectivity under varying failure scenarios; and the \textit{percolation threshold}, identifies the critical point at which the network transitions from being fully connected to disjoint. Such metrics provide valuable  insights into a network's ability to preserve structural integrity under stress or other network disruptions. 
In addition, the \textit{clustering-based} methods such as transitivity, edge clustering coefficient, and modularity evaluate how effectively nodes form cohesive communities. These approaches highlight the network’s capacity to establish and maintain meaningful substructures, thereby reflecting  functional robustness through community dynamics~\cite{heer2020maximising}.

% Finally, there exist \textit{clustering-based} approaches, such as transitivity, edge clustering coefficient, and modularity, which identify interconnectedness in terms of the ability of nodes to form tightly-knight communities. These measures capture the network's ability to form and sustain meaningful substructures, highlighting its functional robustness through community dynamics~\cite{heer2020maximising}.

The existing algorithms focus on SHS detection without specifically addressing the robustness attribute, leaving an important research gap in identifying nodes that not only bridge structural holes but also maintain  resilience in the face of network disruptions. 

\section{Discussion}
To the best of our knowledge, this work is the first to introduce the concept of robust spanners (RS) in large-scale networks. Given the strategic importance of RS nodes in maintaining connectivity and resilience, their applicability could be further explored in the design of adaptive and fault-tolerant systems for emerging domains such as the Internet of Things (IoT) and smart cities.
While the proposed RS identification method demonstrates promising performance, it has the following limitations.
\begin{itemize}
    \item Our current model assumes that the community membership of vertices remains unchanged throughout the entire duration. However, significant topological changes may alter community structures, requiring RSI updates. We aim to develop algorithms for dynamic networks that update RS without full recomputation.
    \item Our method does not support overlapping communities, which we plan to address in future work.
    \item Our current approach relies on a single GPU, which becomes inadequate for very large networks due to memory and performance bottlenecks. A promising, albeit non-trivial, future direction is to develop a distributed multi-GPU extension of the algorithm.
\end{itemize}

%\vspace{-0.1in}
\section{Conclusion}
\label{sec:conclusion}
%\vspace{-0.05in}
% This paper introduced the concept of robust spanners (RS) and proposed a novel scoring mechanism, termed \textit{robust spanning index} (RSI), aimed at identifying potential spanner vertices within large networks. Evaluations of real-world social networks demonstrate that the high-scoring vertices have spanning capacity comparable to those identified by benchmark spanner detection algorithms while demonstrating superior robustness to node and link failures. We detail a parallel algorithm for RS detection and introduce strategies for optimizing memory use and computational effort. We also provide a CUDA-based implementation of this algorithm targeting Nvidia GPUs. This implementation significantly outperforms existing state-of-the-art algorithms in execution speed and is instrumental in RS detection in large-scale social networks prone to evolution and failure.
This work introduces the idea of \textit{robust spanners} (RS) and proposes a novel metric, the \textit{robust spanning index} (RSI), designed to identify potential spanners in large-scale networks. Analysis of real-world social networks reveals that nodes with high RSI scores exhibit spanning capabilities comparable to those identified by state-of-the-art spanner detection methods, while demonstrating better robustness to both node and edge disruptions. We propose a parallel algorithm for identifying RS and introduce optimizations to reduce computational overhead. Additionally, we provide a CUDA-based implementation of this algorithm tailored for Nvidia GPUs.
Our implementation significantly outperforms baseline methods in execution speed and plays a key role in detecting RS in large social networks.

\vspace{-0.1in}
\section*{Acknowledgments}

% \textcolor{red}{You're submitting for the first time, why thanking the reviewers now?} The authors are grateful to anonymous reviewers and associate editor for constructive suggestions that helped significantly improve the quality of the manuscript.
This work was supported by NSF grant awards
ECCS-2319995, OAC-2104078, and CNS-2030624 and OAC-1725755.

\bibliographystyle{IEEEtran}
\bibliography{IEEEabrv,all}

\vspace{-0.4in}
\begin{IEEEbiography}
[{\includegraphics[width=1in,height=1.2in,clip,keepaspectratio]{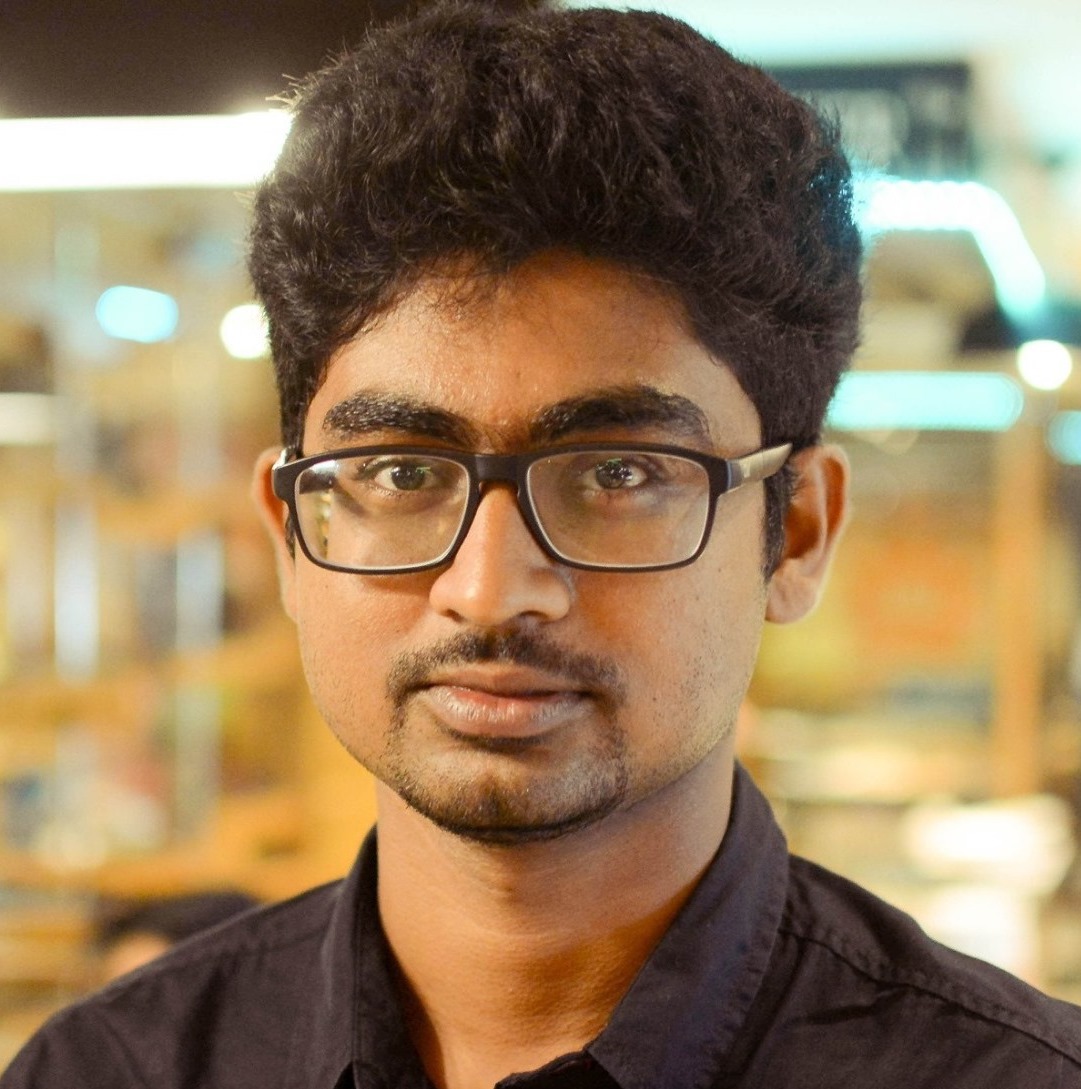}}]{Arindam Khanda} is a postdoctoral researcher in the Department of Computer Science at Missouri University of Science and Technology, where he also earned his Ph.D. in Computer Science. He holds a B.Tech. in Electronics and Communication Engineering from the Institute of Engineering and Management, Kolkata (2015), and an M.Tech. in Software Systems from BITS Pilani (2019).
His research interests include parallel programming models, dynamic graphs, and HPC.
\end{IEEEbiography}

\vspace{-0.5in}
\begin{IEEEbiography}
[{\includegraphics[width=1in,height=1.2in,clip,keepaspectratio]{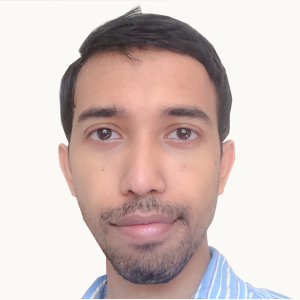}}]{Satyaki Roy} is an Assistant Professor in the Department of Mathematical Sciences at the University of Alabama in Huntsville. His research interests include the design of computational models to analyze complex social and biological systems using statistical and machine learning models, network theory, epidemiology, and bioinformatics. 
\end{IEEEbiography}

\vspace{-0.6in}
\begin{IEEEbiography}
[{\includegraphics[width=1in,height=1.2in,clip,keepaspectratio]{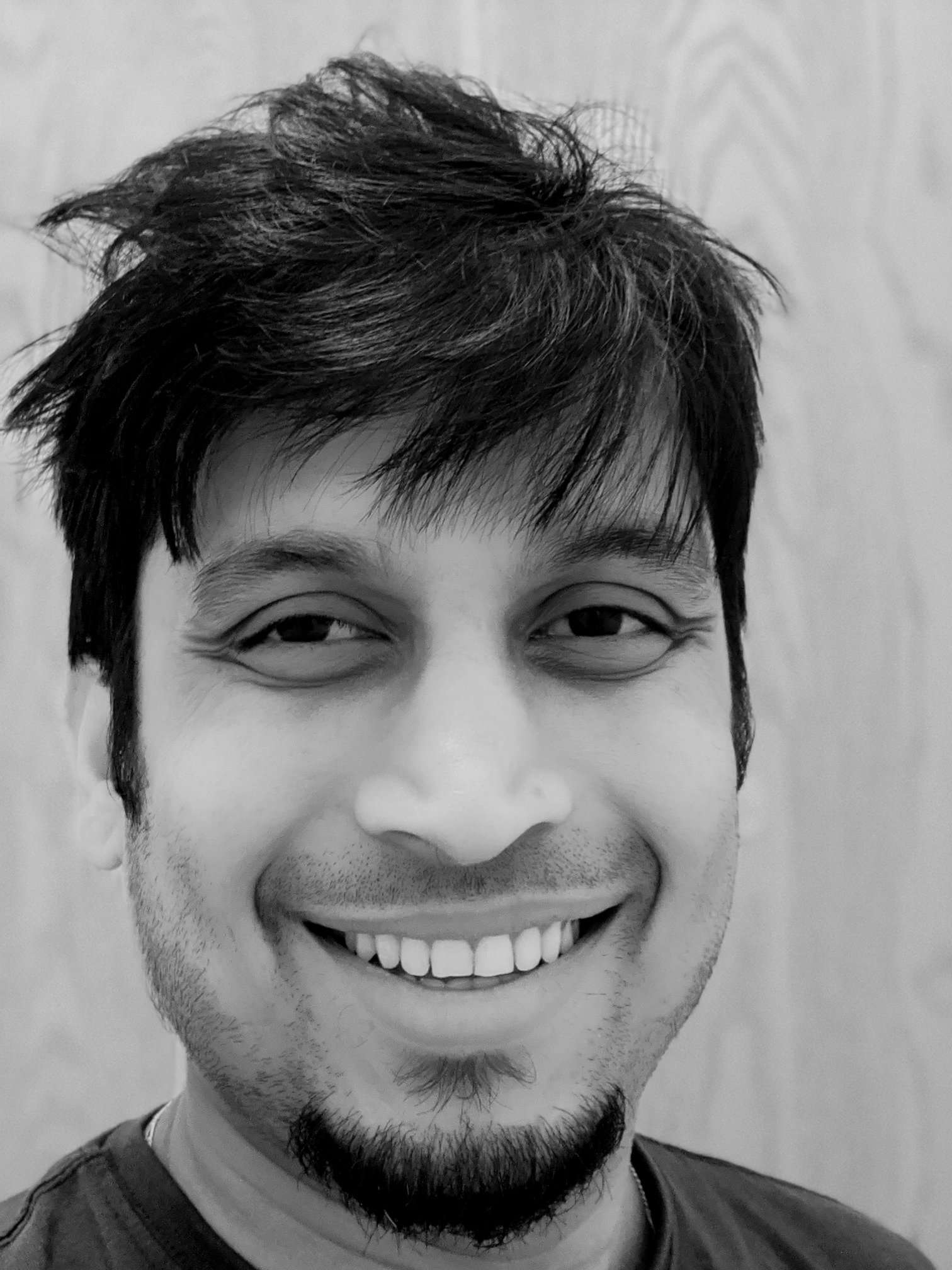}}]{{Prithwiraj} Roy} is the lead data scientist at Global Action Alliance, Inc. He received his Ph.D. in Computer Science from Missouri University of Science and Technology, Rolla, MO, USA. His research interest includes smart grid security, artificial intelligence in cyber-physical systems, and information propagation in complex networks using statistical and machine learning models.
\end{IEEEbiography}

\vspace{-0.5in}
\begin{IEEEbiography}[{\includegraphics[width=1in,height=1.2in,clip,keepaspectratio]{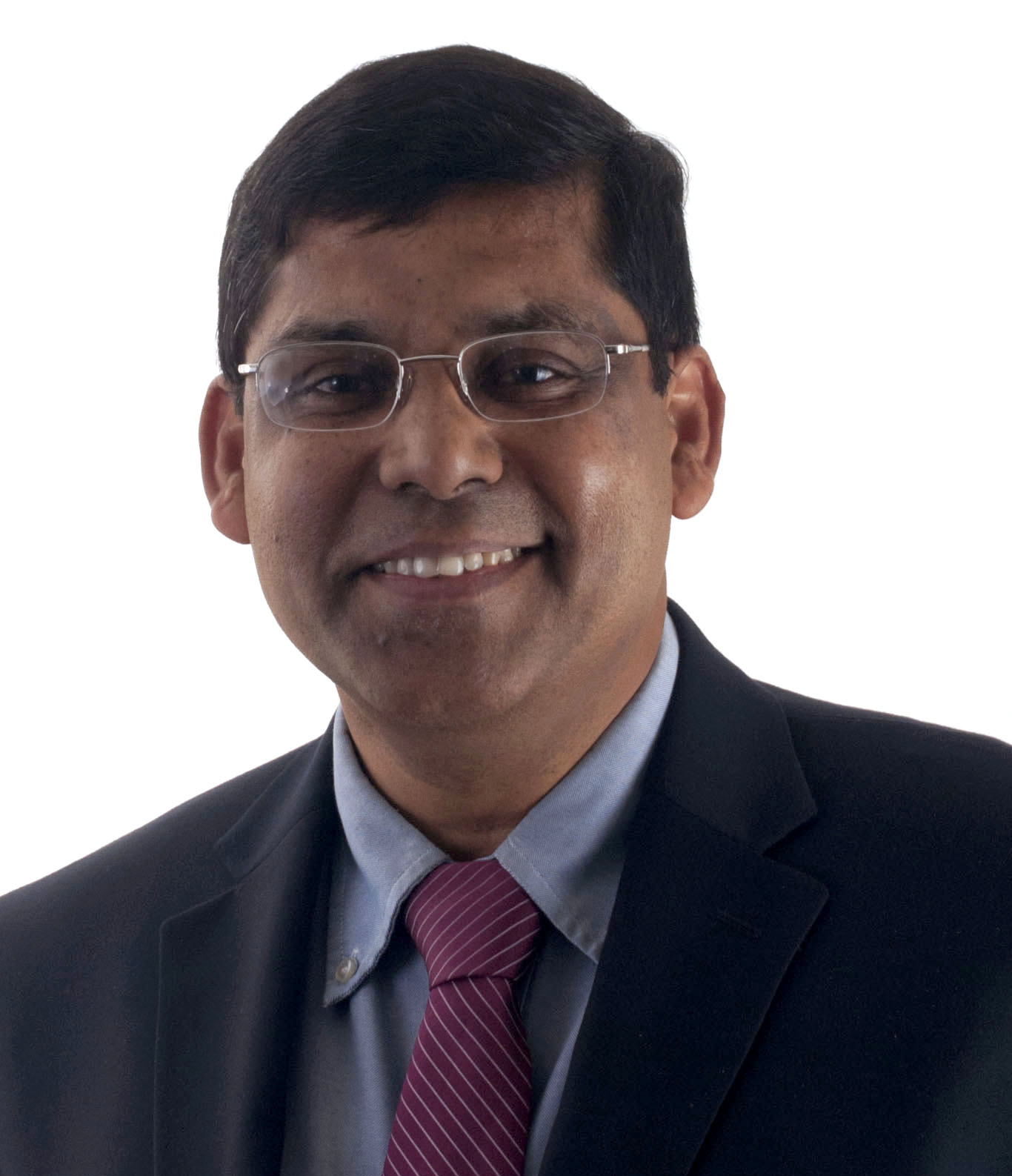}}]{Sajal K. Das}
(Fellow, IEEE) is a Curators' Distinguished Professor of computer science, and Daniel St. Clair Endowed Chair at Missouri University of Science and Technology. His research interests include parallel computing, cloud and
edge computing, graph algorithms, sensor and IoT networks, mobile and pervasive computing, cyber-physical systems,
smart environments, cyber-security, and biological
and social networks. He is the editor-in-chief of Pervasive and Mobile Computing journal, and associate editor of IEEE Transactions on Sustainable Computing, IEEE Transactions on Dependable and Secure Computing, IEEE/ACM Transaction on Networking, and ACM Transactions on Sensor Networks.
\end{IEEEbiography}

\vfill

\end{document}